\documentclass[12pt]{article}

\usepackage[dvips]{graphicx}
\usepackage{epsfig}
\usepackage{amsmath,amsfonts,amssymb,amsthm}
\usepackage{verbatim}
\usepackage{psfrag}
\usepackage{bm}
\usepackage{bbm}
\usepackage[square,comma,sort&compress,numbers]{natbib}
\usepackage{color}
\usepackage{slashed}
\usepackage{soul}
\usepackage{enumitem}
\usepackage[colorlinks=true,urlcolor=blue,anchorcolor=blue,citecolor=blue,filecolor=blue
,linkcolor=blue,menucolor=blue,linktocpage=true,pdfproducer=medialab,pdfa=true
]{hyperref}

\usepackage{dsfont}

\usepackage{epsf,epsfig}
\usepackage{graphics}

\newcommand{\ii}{{\rm i}}
\newcommand{\e}{{\rm e}}
\newcommand{\D}{{\cal D}}
\newcommand{\A}{{\cal A}}

\usepackage[dvipsnames]{xcolor}

\begin{document}
\thispagestyle{empty}

\begin{flushright}
{
\small
KCL-PH-TH/2024-13\\
TUM-HEP-1499/24\\
MITP-24-031
}
\end{flushright}

\vspace{0.4cm}
\begin{center}
\Large\bf\boldmath
The QCD $\theta$-parameter in canonical quantization
%in finite spacetime volume
%WY: although the previous title is more general and perhaps more catching, I do not think it is precise. I do not know how the conclusions will be changed if we consider Euclidean spacetime S^3\times S^1.
\unboldmath
\end{center}

\vspace{0.4cm}

\begin{center}
{Wen-Yuan Ai,$^a$ Björn Garbrecht$^b$ and Carlos Tamarit$^c$}\\
\vskip0.3cm
{\it $^a$Theoretical Particle Physics and Cosmology, King’s College London,\\ Strand, London WC2R 2LS, UK\\[2mm]
 $^b$Physik-Department T70, Technische Universit\"at M\"unchen,\\
James-Franck-Stra{\ss}e, 85748 Garching, Germany\\[2mm]
$^c$PRISMA$^+$ Cluster of Excellence \& Mainz Institute for Theoretical Physics,\\
 Johannes Gutenberg-Universit\"{a}t Mainz, 55099 Mainz, Germany}

\vskip1.4cm
\end{center}

\begin{abstract}

The role of the QCD $\theta$-parameter is investigated in pure Yang--Mills theory in the spacetime given by the four-dimensional Euclidean torus. While in this setting the introduction of possibly unphysical boundary conditions is avoided, it must be specified how the sum over the topological sectors is to be carried out. To connect with observables in real time, we perceive the partition function as the trace over the canonical density matrix. The system then corresponds to one of a finite temperature on a spatial three-torus. Carrying out the trace operation requires canonical quantization and gauge fixing. Fixing the gauge  and demanding that the Hermiticity of the Hamiltonian is maintained leads to a restriction of the Hilbert space of physical wave functionals that generalizes the constraints derived from imposing Gau{\ss}' law. Consequently, we find that the states in the Hilbert space are properly normalizable under an inner product that integrates over each physical configuration represented by the gauge potential one time and one time only. The observables derived from the constrained Hilbert space do not violate charge-parity symmetry. We note that an exact hidden symmetry of the theory that is present for arbitrary values of $\theta$ in the Hamiltonian is effectively promoted to parity conservation in this constrained space. These results, derived on a torus in order to avoid the introduction of boundary conditions, also carry over to Minkowski spacetime when taking account of all possible gauge transformations.

\end{abstract}

\newpage

\hrule
\tableofcontents
\vskip.85cm
\hrule

\section{Introduction}

The present paper concerns the topological term in Yang--Mills theory and the possibility of its observable effects~\cite{Jackiw:1976pf,Callan:1976je,Callan:1977gz} when the theory is {canonically} quantized. As a pure gauge theory, it is specified by the following Euclidean Lagrangian
\begin{align}
\label{QCDlagrangian}
{\cal L}=\frac{1}{2g^2}{\rm tr}F_{\mu\nu}F_{\mu\nu}
-
\frac{\ii}{16\pi^2}\theta\,{\rm tr} F_{\mu\nu} \widetilde F_{\mu\nu}
\,,
\end{align}
where $F_{\mu\nu}=F_{\mu\nu}^a T^a$ with $F_{\mu\nu}^a\equiv \partial_\mu A^a_\nu-\partial_\nu A_\mu^a+f^{abc}A^b_\mu A^c_\nu$ is the field strength tensor and $\widetilde{F}_{\mu\nu}\equiv \frac{1}{2} \varepsilon_{\mu\nu\alpha\beta} F_{\alpha\beta}$ ($\varepsilon_{1234}=1$) is the Hodge dual of $F_{\mu\nu}$. We use the convention ${\rm Tr} (T^a T^b)=\delta ^{ab}/2$, $[T^a,T^b]=\ii f^{abc}T^c$ for the Lie algebra generators $T^a$ and the structure constants $f^{abc}$, and we consider ${\rm SU}(2)$ for simplicity so that $T^a=\tau^a/2$ with $\tau^a$ being the Pauli matrices. The last term in Eq.~(\ref{QCDlagrangian}) is odd under charge-parity ($CP$) conjugation. Through the remainder of the present text, we colloquially refer to Yang--Mills pure gauge theory with the gauge group ${\rm SU}(2)$ as quantum chromodynamics (QCD).

Before outlining matters of canonical quantization, we recall for comparison some salient points about the theory~(\ref{QCDlagrangian}) when evaluated as a path integral in an infinite spacetime volume $\Omega$. The partition function is then given by
\begin{align}
\label{partition:function}
Z=
\int{\cal D}A\,
{\rm e}^{-\lim\limits_{\Omega\to\infty}\int_\Omega {\rm d}^4x\,{\cal L}}\,.
\end{align}
There are a priori no boundary conditions imposed on the path integral. However, configurations of finite action are classified into different topological sectors characterized by integer $\Delta n$---the winding number. We refer to this as topological quantization. The integer $\Delta n$ comes as a consequence of {the limit} $\Omega\to\infty$. This limit must therefore be taken before the sum over the sectors $\Delta n$ is carried out. One then finds that there is no $CP$ violation in the correlation functions~\cite{Ai:2020ptm}. {Furthermore}, leaving the boundary conditions in infinite spacetime unspecified allows the contours of path integration for different $\Delta n$ to be connected by configurations of infinite action so that there exists a full integration contour over all $\Delta n$ in the sense of the Cauchy theorem. {For the individual sectors $\Delta n$, the contours can be constructed using steepest descent methods~\cite{Ai:2022htq}.}

These points stand in opposition to the main share of literature on this topic (see e.g. Refs.~\cite{Callan:1977gz} and~\cite{Yu:2023gdq} for a seminal paper and a recent review) in which the limits are taken the other way around and boundary conditions corresponding to field operator eigenstates that are pure gauge configurations are imposed on finite surfaces. These surfaces are only taken to infinity after the sum over $\Delta n$ is carried out. In that approach, topological quantization does not follow from the definition of the partition function~(\ref{partition:function}) but from the boundary conditions that are imposed ad hoc. {Moreover}, the field configurations with fixed boundary conditions on finite surfaces but different $\Delta n$ cannot be continuously connected through segments of a contour in the gauge field configuration space which do not contribute to the path integral and arise from a nonsingular deformation of the original integration contour. Therefore, for these boundary conditions, there is no integration contour that would allow the application of the Cauchy theorem to the path integral. Calculationally, one finds in that order of limits, i.e. $\Omega\to \infty$ after summation over $\Delta n$, there is $CP$ violation in the correlation functions. For other recent works that question the existence of $CP$ violation in QCD, see for example Refs.~\cite{Nakamura:2021meh,Yamanaka:2022bfj,Torrieri:2020nin}.

Given this context, it is therefore of interest to understand whether the vanishing of $CP$ violation persists in systems that can be quantized in finite volume without imposing unphysical boundary conditions. These can be avoided by choosing manifolds without boundary. The apparently simplest case is to take the three-torus $T^3$ for the spatial hypersurfaces on which canonical quantization is to be carried out. For path integral quantization, we need to specify the range of the temporal coordinate $x_4$ (in the notation used for Euclidean time). In order to project on the ground state, one should take the range of $x_4$ to infinity. This way, we do not need to specify boundary conditions on the spatial hypersurfaces $T^3$ at $x_4\to\pm\infty$. However, then we are back to the discussion on the partition function~(\ref{partition:function}) in infinite spacetime volumes~\cite{Ai:2020ptm}.

It is therefore of interest to compactify time as well. A way to achieve this without specifying ad hoc boundary conditions is to impose periodicity in $x_4$ so that the spacetime becomes $T^4$. Such a setup is physical as the resulting partition function corresponds to the one of a canonical ensemble kept at a temperature $T$ where the range of $x_4$ corresponds to $1/T$. Now, for the theory~(\ref{QCDlagrangian}) on $T^4$ once again the field configurations decompose into equivalence classes of integer $\Delta n$. While this is a well-known fact, we review the reasons for this topological quantization in Section~\ref{sec:topology:T4} in order to relate it to the topological aspects that come into play in the subsequent discussion of canonical quantization, a central subject of the present paper.

To compute observables at finite temperature, we need an expression for the partition function that corresponds to Eq.~(\ref{partition:function}) but now for $T^4$ instead of an infinite $\Omega$. The answer is not as straightforward as it may seem. On $T^4$, there are no continuous deformations that connect sectors of different $\Delta n$ and consequently, these do not lie on a single Cauchy contour. Furthermore, fixing the winding number $\Delta n$ o $T^4$ leads to a well-defined Euclidean quantum field theory for each $\Delta n$ so that there is a priori no necessity for summing over $\Delta n$. We therefore need a more detailed reasoning to determine whether we need to sum over the topological sectors and how we must weigh the contributions in such a sum, in particular in terms of phase factors.

A unique way of specifying the finite-temperature partition function is to refer to its definition as
\begin{align}
\label{eq:partition_function}
Z={\rm tr}\,{\rm e}^{-H/T}\,,
\end{align}
where $H$ is the Hamiltonian and the trace is over the physical states in the Hilbert space. We thus need to gather enough facts about the Hilbert space in order to derive the partition function. In particular, to carry out the trace operation, we must establish in what sense the states are normalizable and that we can fix the gauge in order not to sum over redundant states. While we are focusing here on canonical quantization, we note that when the observables obtained in this approach are independent of the particular gauge fixing, one can derive an expression for the path integral in which the gauge symmetry along with Lorentz invariance becomes manifest again~\cite{Fradkin:2021zbi}. Upon evaluating path integrals, gauge fixing is then in practice reintroduced in a different way using the standard functional methods, {i.e. the Faddeev–-Popov procedure.}

To identify the Hamiltonian and the Hilbert space, we review some of the basic facts on canonical quantization of QCD theory in Section~\ref{sec:canonical}. In Section~\ref{sec:gauge:fixing}, we specify in what sense states should be normalizable after gauge fixing and make provisions in our notation for excluding fluctuations and derivatives from the Hamiltonian in the direction of gauge redundancies.

A particular set of gauge transformations that we discuss in Section~\ref{sec:transition:function} are those that change the transition functions. The latter are required to construct the maps from $T^3$ into the vector potential  $\mathbf A$ in the temporal gauge. When we fix these transition functions, the remaining gauge transformations must be periodic over $T^3$  and hence decompose again into integer equivalence classes. In particular, those periodic transformations that are not continuously connected with the identity are referred to as large gauge transformations. The set of large gauge transformations (corresponding to $n\neq 0$ below) can be generally expressed as 
\begin{align}
\label{eq:large-transf}
\{G_n\}=Z_n\otimes \{G_0\}\,,
\end{align}
where $Z_n$ is the additive group of integers and $\{G_0\}$ is the set of so-called small gauge transformations that are connected with the identity. 
At this point, one can introduce wave functionals that are eigenstates of large gauge transformations $G_n$ with $\exp(\ii n \theta^{(i)}$)  as their eigenvalues. The superscript here allows to write expressions involving wave functionals with different eigenvalues $\exp(\ii n \theta^{(j)})$ with  $\theta^{(i)}\not=\theta^{(j)}$ and also indicates that these eigenvalues are in the first place distinct from $\theta$ in the Lagrangian~(\ref{QCDlagrangian}). 

However, we note that the states for different $\theta^{(i)}$ based on a gauge-fixed inner product are not orthogonal. In turn, when relaxing the gauge fixing in the inner product we encounter the problem noted in Ref.~\cite{Okubo:1992na} that the states are not properly normalizable. To resolve these issues, we pursue the direction that the Hilbert space with different $\theta^{(i)}$ is too large. The correct value of $\theta^{(i)}$ is uniquely determined by demanding that $(i)$ one can fix the gauge; $(ii)$ the operators corresponding to observables remain Hermitian; and $(iii)$ the observables are the same in different gauges. In Section~\ref{sec:quantum:constraints}, we derive the constraints on the wave functionals that follow from these requirements. As a familiar example on which to apply this reasoning, in Section~\ref{sec:Gauss} we review that imposing Gau{\ss}' law restricts the Hilbert space to states that are gauge singlets with respect to transformations that leave the transition functions invariant. In turn, we can use the requirement of gauge invariance to restrict the Hilbert space to states that observe Gau{\ss}' law.

In Section~\ref{sec:first:quantize:then:constrain} we make the analogous argument for transformations changing the transition function. As it turns out, this indeed determines the value of $\theta^{(i)}$ so that the gauge-fixed inner product can be used without problem on the remaining Hilbert space. The allowed value for $\theta^{(i)}$ is the one that complements the $CP$-odd phase from the Hamiltonian so that there remain no $CP$-violating effects. {While the results to that point have been derived on a torus in order to avoid dealing with boundary conditions, in Section~\ref{sec:Minkowski}, we show that the same reasoning applies to Minkowski spacetime as well when one does not restrict  gauge {transformations} at spatial infinity, and we also show how the results for the wave functionals explain $CP$ conservation in amplitudes evaluated over finite time intervals without periodic temporal boundary conditions. In Section~\ref{sec:Parity}, a new discrete symmetry of the Hamiltonian operator is identified. It can be used to show that for the gauge fixed Hamiltonian, energy eigenstates in the physical Hermitian subspace can always be constructed as eigenstates of the usual parity. Conclusions can be found in Section~\ref{sec:conclusions}.}

\section{Topological quantization on the four-torus}
\label{sec:topology:T4}

While it is not necessary to consider $T^4$ in order to show $CP$ conservation in canonical quantization on the spatial $T^3$, the canonical density matrix in its path integral representation on $T^4$ is an object of special interest. {In that context, it is essential that} the winding number $\Delta n$ is an integer on $T^4$, i.e.
\begin{align}
\label{topo:quant:torus}
\Delta n=\frac{1}{16\pi^2}\int_{T^4} {\rm d}^4 x F \widetilde F\in\mathbbm{Z}\,.
\end{align}
Proofs of the topological quantization~(\ref{topo:quant:torus}) can be found e.g. in Ref.~\cite{Leutwyler:1992yt}. In order to see in what way topology on $T^3$ and $T^4$ are related, we show here explicitly how Eq.~(\ref{topo:quant:torus}) holds.

To that end, we recall that the integrand can be written as a total divergence,
\begin{align}
\label{total:divergence}
\frac{1}{16\pi^2}{\rm tr}F_{\mu\nu}\widetilde F_{\mu\nu}=\partial_\mu K_\mu\,,
\end{align}
where
\begin{align}
\label{K:mu}
K_\mu=\frac{1}{4\pi^2}\epsilon_{\mu\nu\alpha\beta}{\rm tr}\left[\frac12 A_\nu\partial_\alpha A_\beta-\frac{\rm  i}{3}A_\nu A_\alpha A_\beta\right]\,.
\end{align}
We recall that for $SU(2)$, ${\rm tr}(T^a T^b T^c)=\ii f^{abc}/4$ so that the above expression can be written in terms of the components of the gauge fields as
\begin{align}
K_\mu=\frac{1}{16\pi^2}\epsilon_{\mu\nu\alpha\beta}\left[A^a_\nu \partial_\alpha A^a_\beta +\frac{1}{3} f^{abc} A^a_\nu A^b_\alpha A^c_\beta\right]\,.
\end{align}

Next, consider the hypercube with edges of length $a_i$, $i=1,\ldots,4$ from which we can obtain $T^4$ by gluing the opposite three-faces together. On the opposite three-faces of the hypercube, the vector potential must match modulo a gauge transformation. For example,
\begin{align}
\label{transition:cube}
A_\mu(\varepsilon,x_2,x_3,x_4)=&U_{1}^{-1}(x_2,x_3,x_4) A_\mu(a_1-\varepsilon,x_2,x_3,x_4) U_{1}(x_2,x_3,x_4)\notag\\
&+{\rm i} U_{1}^{-1}(x_2,x_3,x_4)\partial_\mu U_{1}(x_2,x_3,x_4)\,,
\end{align}
and accordingly for the remaining pairs of three-faces. Here, $\varepsilon$ is an infinitesimal length that we have introduced in order to formulate this matching condition.

Equation~(\ref{transition:cube}) is routinely imposed when putting gauge fields on a torus~\cite{tHooft:1979rtg} and does not appear to require more explanation than given above. When we insist on stating it in a bit more technical terms, we can take an open cover $\{{\cal U}_\alpha\}$ of $T^4$. Gauge-covariant quantities on $T^4$ can then be defined with the help of transition functions $t_{\alpha\beta}:{\cal U}_\alpha\cap {\cal U}_\beta\to {\rm SU}(2)$. In particular, for $x\in {\cal U}_\alpha\cap {\cal U}_\beta$, there is the compatibility condition
\begin{align}
A_\mu^\beta=t_{\alpha\beta}^{-1} A_\mu^\alpha t_{\alpha\beta} +{\rm i} t_{\alpha\beta}^{-1}\partial_\mu t_{\alpha\beta}\,,
\end{align}
{where $A^{\alpha,\beta}_\mu$ are the gauge potentials defined in the respective ${\cal U}_{\alpha,\beta}$.}
For consistency, the transition functions have to satisfy the cocycle condition: For $x\in {\cal U}_\alpha \cap {\cal U}_\beta \cap {\cal U}_\gamma$ it must hold that $t_{\alpha\gamma}=t_{\alpha\beta} t_{\beta\gamma}$.

\begin{figure}[t]
\centering
\includegraphics[scale=0.5]{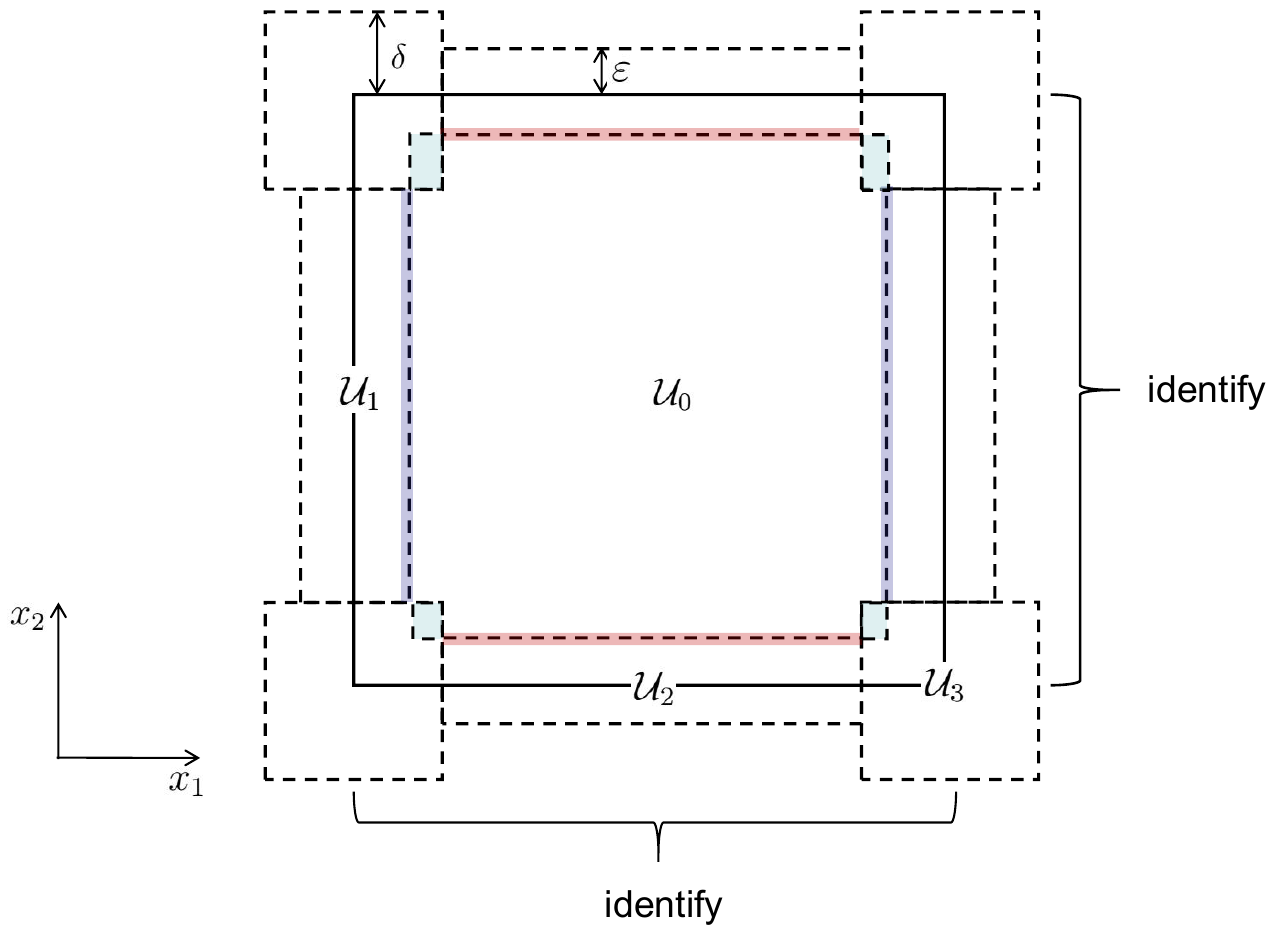}
\caption{Illustration of a cover of $T^2$ (the region enclosed by the solid line) using four open sets (regions enclosed by dashed lines and denoted as ${\cal U}_\alpha$ with $\alpha=0,1,2,3$) of which overlapping regions (shaded regions in color) have a minimal area when $\delta,\varepsilon\rightarrow0$ (keeping $\delta>\varepsilon$ in taking the limit)~\cite{vanBaal:1982ag}.}
\label{fig:T2}
\end{figure}

To arrive at Eq.~(\ref{transition:cube}), it is advantageous to reduce the overlapping regions of the open sets ${\cal U}_\alpha$ to a minimal ``volume''~\cite{vanBaal:1982ag}. For $T^n$ the minimal number of the open sets for such a covering is $2^n$. One can illustrate this for $T^2$ as an example, as shown in Figure~\ref{fig:T2}. Note the identification for opposite edges and faces so that the four squares at the corners are identified as one single open set ${\cal U}_3$, the opposite rectangles are identified as ${\cal U}_1$ and ${\cal U}_2$ respectively. The respective overlaps between ${\cal U}_0$ and ${\cal U}_\alpha$, $1\leq\alpha\leq 3$ (e.g. the two lines in blue for $\alpha=1$) have disconnected parts. From the figure, one can see that $\delta>\varepsilon$ but eventually, one should take $\delta,\varepsilon\rightarrow 0$ to have a minimal area for the overlapping regions.

For $T^4$, we take ${\cal U}_0$ as the open set that covers the hypercube just without its boundaries. Fifteen more sets ${\cal U}_1,\ldots,{\cal U}_{15}$ can be arranged along the cells (three-faces), faces (two-faces), edges (one-faces) and vertices (zero-faces) of the hypercube so that one obtains a complete cover. For definiteness, let the ${\cal U}_i$ for $i=1,\ldots,4$ be the sets covering the cells with $x_i=0$ along with their opposites $x_i=a_i$ up to the infinitesimal shell along the faces {(analogues of ${\cal U}_{1,2}$ in the $T^2$ example in Figure~\ref{fig:T2}}). Then shrink  ${\cal U}_1,\ldots,{\cal U}_{15}$ so that they cover just infinitesimal neighbourhoods of the cells, faces, edges and vertices. Analogous to the case of $T^2$ the identification of $x_i$ with $x_i+a_i$ determines the
number of open sets needed to cover the full $T^4$. Note that in this construction, the intersections ${\cal U}_0 \cap {\cal U}_\alpha$, $1\leq\alpha\leq 15$ are not connected, as illustrated on the example of $T^2$. The transition functions over each pair of opposite cells of the hypercube can then be composed from the transition functions on the disjoint subdomains according to e.g. 
\begin{align}
    U_1(x_2,x_3,x_4)=t_{10}(\varepsilon,x_2,x_3,x_4) t_{01}(a_1-\varepsilon,x_2,x_3,x_4)
\end{align}
so that Eq.~(\ref{transition:cube}) follows.

Then we choose a gauge so that the gauge potential is periodic in the { three spatial directions}~\cite{Luscher:1982ma}, i.e., $A_\mu(x+a_i \hat e_i)=A_\mu(x)$ {for $i=1,2,3$}. This can be accomplished because we do not impose boundary conditions on the allowed gauge transformations on the three-faces at $x_4=0,a_4$. As a consequence of this gauge choice, the net flux of $K_\mu$ through the opposite three-faces $x_{1,2,3}=0$ and $x_{1,2,3}=a_{1,2,3}$ vanishes. In addition, we can impose temporal gauge so that
\begin{align}
\label{temporal:gauge}
A_0=0
\end{align}
everywhere.

Now consider the opposite three-faces at $x_4=0$ and $x_4=a_4$. Given the gauge choices made thus far, all of the total flux of $K_\mu$ through the surface of the hypercube must go through these two three-faces. Let
\begin{align}
F_4= [0,a_1]\times[0,a_2]\times[0,a_3]
\end{align}
be the set pertaining to these three-faces on which $U_4$ is defined corresponding to Eq.~(\ref{transition:cube}). We can conclude that $U_4(\mathbf y)=\mathbbm 1$ for $\mathbf y\in \partial F_4$ because of the periodicity of the gauge potential in the spatial directions that is assumed here. Therefore, we can identify the points in $\partial F_4$ so that in this sense, the three-face is homeomorphic with $S^3$. Now given the winding number
\begin{align}\label{eq:nu}
\nu(U;S)=\frac{1}{4\pi^2}\varepsilon_{ijk}\int_{S}{\rm d}^3x\, \frac 16 {\rm tr}\left[ U^{-1} (\partial_i U) U^{-1}(\partial_j U) U^{-1} (\partial_k U)\right]\,,
\end{align}
the transition functions $U_4(\mathbf x)$ therefore constitute homotopy classes with integer winding number, i.e. $\nu(U_4;F_4)\in\mathbbm Z$ for $U_4=\text{constant}$ on $\partial F_4$.

It remains to be shown that this implies Eq.~(\ref{topo:quant:torus}), in particular when the field configurations on the three-face $F_4$ are not a pure gauge. Consider the Chern--Simons functional
\begin{align}
\label{CS:functional}
W[\mathbf A]=\frac{1}{4\pi^2} \varepsilon_{ijk}\int_S {\rm d}^3x\,{\rm tr}\left[\frac12 A_i\partial_j A_k-\frac{\rm i}{3} A_i A_j A_k\right]
\end{align}
over some three-dimensional hypersurface $S$. Now for a gauge transformation on $S$
\begin{align}
\label{gaugetrafo:spatial}
\mathbf A^\prime=U \mathbf A U^{-1} + {\rm i} U^{-1} \nabla U
\end{align}
the difference between the Chern--Simons functionals is~\cite{Luscher:1982ma}
\begin{align}
\label{CS:diff}
W[\mathbf A^\prime]-W[\mathbf A]=\nu(U;S)-\frac{{\rm i}}{8\pi^2}\varepsilon_{ijk}\int_{S}{\rm d}^3x\,\partial_j{\rm tr}\left[A_k(\partial_i U) U^{-1}\right]\,.
\end{align}
Substituting $U=U_4$ and $S=F_4$, this is the difference between the Chern--Simons numbers that flows through the three-faces at $x_4=0$ and $x_4=a_4$. As we have made use of the gauge freedom so that $U_4$ is constant on $\partial F_4$ and that $\mathbf A$ is periodic on $F_4$, ${\rm tr} [A_k (\partial_i U_4)U^{-1}_4]$ is periodic on $F_4$ and the integral vanishes by Gauß' theorem. Furthermore, the Chern--Simons flux through the remaining faces vanishes and $\nu(U_4,F_4)\in \mathbbm Z$. Hence, using Eq.~(\ref{total:divergence}), we recognize that $\Delta n=W[\mathbf A^\prime]-W[\mathbf A]\in\mathbbm Z$, i.e. Eq.~(\ref{topo:quant:torus}) holds. Since Eq.~(\ref{topo:quant:torus}) is manifestly gauge invariant, we can change the above construction to general gauges without changing this result.

\section{Canonical quantization of gauge theory}
\label{sec:canonical}

Canonical quantization of non-Abelian gauge theory is reviewed in e.g. Ref.~\cite{Jackiw:1979ur} which includes many details. This section contains a summary of the main points that are relevant in the present context.

To this end, we start by casting the Euclidean Lagrangian~(\ref{QCDlagrangian}) to Minkowski spacetime
\begin{align}
\label{eq:Lagrangian}
{\cal L}=&-\frac1{2 g^2} {\rm tr} F_{\mu\nu} F^{\mu\nu}
+\frac{1}{16\pi^2}\theta\,{\rm tr} F_{\mu\nu} \widetilde F^{\mu\nu}\,.
\end{align}
In the following, we make occasional use of the covariant derivative
\begin{align}
D_\mu=\partial_\mu-{\rm i}  A_\mu^a\frac{\tau^a}{2}\,.
\end{align}
In terms of this operator,
the field strength tensor
\begin{align}
F_{\mu\nu}= F_{\mu\nu}^a \frac{\tau^a}{2}
\end{align}
can be expressed as
\begin{align}
[D_\mu,D_\nu]=-{\rm i} F_{\mu\nu}\,.
\end{align}

In canonical quantization, as long as we are not interested in the time evolution, we are concerned with $T^3$ instead of $T^4$. However, the discussion in the previous section is related to the present context because we can, without loss of generality, take $F_4$ as the spatial hypersurface $T^3$ that the quantum field is defined on. And $U_4$ can {here be identified} as a gauge transformation for the gauge field on the same spatial hypersurface $T^3$.

The most straightforward approach to canonical quantization of Yang--Mills theory appears to go through the temporal gauge~(\ref{temporal:gauge}), in which the non-Abelian electric and magnetic fields are given by
\begin{align}
{g}\mathbf{E}^a=&{-}\frac{\partial}{\partial t}\,\mathbf A^a\,,\\
{g} \mathbf B^a=&\nabla\times \mathbf A^a-\frac{1}{2}\,f^{abc} \mathbf{A}^a \times\mathbf{A}^b\,.
\label{B:field}
\end{align}
In terms of these, the canonical momentum conjugate to $\mathbf A^a$ is
\begin{align}
\label{eq:Pi-E-B}
{g}{ \mathbf{\Pi}^a=g\frac{\partial{\cal L}}{\partial \dot{\mathbf A}^a}={-} \mathbf E^a+\frac{g^2}{8\pi^2} \theta \mathbf B^a}\,.
\end{align}
The coordinates $\mathbf A^a$, $\mathbf\Pi^a$ are canonical and satisfy the corresponding Poisson brackets. This is the main advantage over imposing more restrictive gauge conditions, where one would need to generalize to Dirac brackets. While this is possible for the Abelian case, such procedure appears to face prohibitive complications for non-Abelian theories. On the other hand, using the above set of canonical coordinates requires extra steps when quantizing the theory. Since the temporal gauge leaves substantial gauge freedom in the form of the transformation~(\ref{gaugetrafo:spatial}), unphysical states and gauge redundancies must be eliminated subsequent to quantization. The present procedure may therefore colloquially be referred to as ``first quantize then constrain" (modulo the constraint~(\ref{temporal:gauge}) that is imposed before quantization). We therefore yet need to specify how to narrow down the Hilbert space to physical states and eventually to fix the gauge completely.

The operators corresponding to the canonical coordinates in temporal gauge must observe the commutation relations: 
\begin{align}
[A^{a,i}(\mathbf x),\Pi^{b,j}(\mathbf x^{\,\prime})]={\rm i} \delta^{ij}\delta^{ab}\delta^3(\mathbf x-\mathbf x^{\,\prime})
\,,
\quad
[\Pi^{a,i}(\mathbf x),\Pi^{b,j}(\mathbf x^{\,\prime})]=0\,.
\label{eq:canonical:commutators}
\end{align}
These commutators hold for
\begin{align}
\label{can:mom:gauge}
\mathbf{\Pi}^a=\frac{\delta}{{\rm i}\delta \mathbf A^a}+\theta_\Pi\frac{g}{8\pi^2} \mathbf{B}^a\,.
\end{align}
Here, $\theta_\Pi$ is an arbitrary constant that is real for $\mathbf \Pi^a$ to be Hermitian. Combining Eqs.~\eqref{eq:Pi-E-B} and~\eqref{can:mom:gauge} yields the operator $\mathbf{E}^a$,
\begin{align}\label{eq:Efield}
\mathbf{E}^a=-\left(g\frac{\delta}{\ii\delta\mathbf{A}^a}-\frac{g^2}{8\pi^2}\left(\theta-\theta_\Pi\right) \mathbf{B}^a\right)\,.
\end{align}
Substituting the above into the Hamiltonian density, one obtains
\begin{align}
\label{Hamiltonian:gauge}
{\cal H}=\frac 12\left[({\mathbf E^a})^2+(\mathbf B^a)^2\right]
=\frac 12 \left[\left(g\frac{\delta}{{\rm i}\delta \mathbf A^a}-\frac{g^2}{8\pi^2} (\theta-\theta_\Pi)\mathbf B^a\right)^2+(\mathbf B^a)^2\right]\,.
\end{align}
The Hamiltonian
\begin{align}
H=\int{\rm d^3}x\,{\cal H}
\end{align}
acts on states represented through wave functionals $\Psi[\mathbf A]$. However, since we have not fixed the gauge prior to quantization, the Hilbert space of all functionals is too large, i.e. it contains unphysical states as well as gauge redundancies. In the following sections, we review how the unphysical states are eliminated.

\section{Gauge redundancies and gauge fixing}
\label{sec:gauge:fixing}

As reviewed in Section~\ref{sec:canonical}, the quantized theory defined through the canonical commutation relations~(\ref{eq:canonical:commutators}) may yet allow for unphysical states and entails gauge redundancies that must be removed by gauge fixing. In the present section, we therefore introduce some notations in order to deal with this matter.

We first get back to the point that there is still some residual gauge freedom. As different gauges are redundant descriptions of the same physical system, it is possible to fix the gauge in such a way that there is a one-to-one correspondence between the functions $\mathbf A(\mathbf x)$ and classical physical  configurations, i.e. that there is a complete gauge fixing. This implies that we can express the inner product of two states as

\begin{align}
\label{inner:product:g.f.}
\langle\Psi|\Phi\rangle=\int_{\cal A}{\cal D}\mathbf A f_{\cal A}[\mathbf A] \Psi^*[\mathbf A] \Phi[\mathbf A]\,,
\end{align}
where the subscript on the integral indicates complete gauge fixing. That is, we integrate over a manifold $\cal A$ on which each gauge inequivalent configuration is being represented one time and one time only. The factor $f_{\cal A}[\mathbf A]$ is generally present in order to maintain the gauge covariance of the fixed integration measure, i.e. the measure ${\cal D }\mathbf A f_{\cal A}[\mathbf A]$ is translation invariant when restricted to ${\cal A}$. For example, when fixing the gauge by inserting a gauge condition as an argument of a delta-function, $f_{\cal A}(\mathbf A)$ would be given by the absolute value of the inverse determinant of the gauge condition. That being said, we could also decide to integrate over redundant configurations if the corresponding integration volume is finite so that we can still properly normalize the states. Nonetheless, it should always be possible if not advantageous to use an inner product as in Eq.~(\ref{inner:product:g.f.}) that does not extend over redundant configurations. {In turn,} noting that Eq.~\eqref{inner:product:g.f.} can represent a physical transition amplitude, it should be gauge invariant, so that extending the integration over {an infinite measure of} gauge redundant configurations leads to unphysical infinite answers. In particular, {were there no gauge fixing}, the inner product of Eq.~\eqref{inner:product:g.f.}  would lead to states of infinite norm, in contradiction with the postulates of quantum mechanics.

The restriction of the Schrödinger equation to a certain subset of field configurations is naturally defined through bases of the field configuration space $\{\mathbf A^a(\mathbf x)\}$. Let $\mathbf G^a(\mathbf x; \sigma)$ be a basis for the configuration space, where $\sigma$ is the multiindex
%(e.g. momentum and polarizations)
that labels the particular basis modes. Then, a derivative in the direction of $\mathbf G^a(\mathbf x; \sigma)$ can be defined as
\begin{align}
\label{eq:trafo:FD}
\frac{\delta}{\delta A(\sigma)}\Psi[\mathbf A]=\int{\rm d}^3 y \sum\limits_{i,a} G^a_i(\mathbf y; \sigma) \frac{\delta}{\delta A^a_i(\mathbf y)}\Psi[\mathbf A]\,.
\end{align}
Note that the basis $G_i^a(\mathbf y; \sigma=\{j,b,\mathbf x\})=\delta^3(\mathbf y-\mathbf x)\delta^{ij}\delta_{ab}$ reproduces the standard derivative $\delta/\delta A_b^j(\mathbf x)$. If the basis is orthonormal, i.e. if the following identity is satisfied,
\begin{align}
 \int\limits_\sigma\hskip-.5cm\scalebox{1.}{$\sum$}\,{G_i^{a}(\mathbf x;\sigma) G_j^{b}(\mathbf y;\sigma)} = \delta^{(3)}(\mathbf{x-y})\delta^{ab}\delta^{ij},
\end{align}
we can conversely write
\begin{align}
\frac{\delta}{\delta A^a_i(\mathbf x)}\Psi[\mathbf A]=\int\limits_\sigma\hskip-.5cm\scalebox{1.}{$\sum$} G^{a}_{i}(\mathbf x; \sigma)\frac{\delta}{\delta A(\sigma)}\Psi[\mathbf A]\,.
\end{align}

When we identify some basis modes ${\mathbf G}^a(\mathbf x; \sigma_{\text{gauge}})$ with infinitesimal gauge transformations of a given configuration $\mathbf A^a(\mathbf x)$, we can fix in the Schr\"odinger equation the gauge with respect to such transformations by omitting the derivatives $\delta/\delta  A(\sigma_{\text{gauge}})$. In the following, we apply this to transformations generated by Gau{\ss}' law, which are periodic over the torus as well as to more general aperiodic transformations that change the transition function.

\section{Fixing the transition function and normalization of physical states}
\label{sec:transition:function}

To derive the Hamiltonian~(\ref{Hamiltonian:gauge}), no extra gauge fixing in addition to the temporal gauge has been applied. This is the assumption that we are going to work with in the subsequent sections. Before going ahead, in this section we consider what happens when we fix the transition function because this way one may arrive at constructions on the three-torus that correspond to the $\theta$-vacua usually discussed for infinite space. While the temporal gauge imposes no restriction on the transition function~(\ref{transition:cube}), let us consider here $U_i=\mathbbm 1$ for $i=1,2,3$. This way, the gauge potential is restricted to be periodic on the three torus and  $U_4=\mathbbm 1$ on $\partial F_4$.

It is essential to note that other choices of the  transition functions are also possible and generally lead to non-periodic gauge potentials $\mathbf A$ on $T^3$ even though the physical configuration remains periodic. The assumption made in the present section, that one can simply ignore field configurations that are not periodic in $\mathbf A$ when restricting the form of the wave functionals, is something that should therefore be verified. Fixing the gauge and thereby the transition function before quantization corresponds to constraints that are difficult to implement in non-Abelian gauge theories. In Section~\ref{sec:quantum:constraints}, we therefore derive how to restrict the Hilbert space in order to allow for gauge fixing subsequent to quantization.

Considering again Eq.~(\ref{CS:diff}) with $S=F_4$, $U=U_4$ and assuming $U_4=\mathbbm 1$ on $\partial F_4$, we recognize that for a gauge transformation~(\ref{gaugetrafo:spatial}) under the imposed constraint of periodic vector potentials, $W[\mathbf A^\prime]-W[\mathbf A]\in\mathbbm Z$. We emphasize that this constraint that leads to topological quantization in this sense is a choice that one does not necessarily have to make. Going ahead notwithstanding this point, for a gauge transformation $U^{(1)}_4$ so that $W[\mathbf A^\prime]-W[\mathbf A]=1$, using the fact that it commutes with the Hamiltonian $H$, one concludes that there are eigenstate wave functionals $\Psi_{\theta^{(i)}}[\mathbf A]$ of $H$ and $U_4^{(1)}$ so that 
\begin{align}
\label{Psi:EV:U4}
\Psi_{\theta^{(i)}}[A_{(U_4^{(1)})^n}] ={\rm e}^{{\rm i}n \theta^{(i)}} \Psi_{\theta^{(i)}}[\mathbf A]
\end{align}
with some real $\theta^{(i)}$. According to the Bloch theorem, these can be written as 
\begin{align}
\Psi_{\theta^{(i)}}[\mathbf A]=\psi_{\theta^{(i)}}[\mathbf A]{\rm e}^{{\rm i}W[\mathbf A]\theta^{(i)}}\,,
\end{align}
with Eq.~(\ref{CS:functional}) for $W[\mathbf A]$, $S=F_4$, and $\psi_{\theta^{(i)}}[\mathbf A]=\psi_{\theta^{(i)}}[\mathbf A^\prime]$. We have introduced here the shorthand notation
\begin{align}
    \mathbf A_U=U \mathbf A U^{-1}+{\rm i} U^{-1}\nabla U\
\end{align}
for a gauge transformed vector potential.

So far, there are three angles that appear in the canonical quantization of QCD: $\theta$ in the Lagrangian, $\theta_\Pi$ in the canonical field momentum operator and $\theta^{(i)}$ in the wave functionals. (When introducing fermions, which we do not consider in the present work, there is an additional angular parameter from the Dirac mass.) The combination $\theta-\theta_\Pi$ should appear in the Lagrangian of the path integral when one constructs the latter from the Hamiltonian~\eqref{Hamiltonian:gauge}. One might think that the freedom to choose $\theta_\Pi$ could be used to remove the theta parameter. However, the path integral should be weighted by wave functionals that arise from the projection from the initial and final physical states onto the gauge configurations (see Eq.~\eqref{eq:Z_Minkowski} below). Therefore, the theta parameter usually discussed in QCD, where one often does not distinguish the additional angular parameters, corresponds to $\theta-\theta_\Pi-\theta^{(i)}$ which is an invariant (because $\theta_\Pi-\theta^{(i)}$ is invariant)~\cite{Jackiw:1979ur}. However, as we shall show, a consistent canonical quantization of the QCD on $T^3$ implies that the spectrum contains only states for which the invariant $\theta-\theta_\Pi-\theta^{(i)}$ vanishes. As a consequence, the path integral formulation for the same theory, i.e. QCD in a finite volume, can only have a vanishing theta parameter. In Section~\ref{sec:Minkowski}, we also discuss how this result carries over to infinite volume, i.e., $\mathbb{R}^3$, when we discard the hypothesis~\eqref{eq:U_constraint}.

Now take two states $\Psi^{(a,b)}_{\theta^{(i,j)}}$ of this type, where $a,b$ are labels for states in orthonormal bases of the subspaces with fixed $\theta^{(i)}$,  $\theta^{(j)}$, respectively. We may then note that
\begin{align}
\int {\cal D}\mathbf A \Psi^{(a)*}_{\theta^{(i)}}[\mathbf A] \Psi^{(b)}_{\theta^{(j)}}[\mathbf A]=&\sum\limits_{\nu=-\infty}^\infty\int_{0\leq W[\mathbf A]< 1}{\cal D}\mathbf A
{\rm e}^{-{\rm i}(\theta^{(i)}-\theta^{(j)})(W[\mathbf A]+\nu)}\psi^{(a)*}_{\theta^{(i)}}[\mathbf A] \psi^{(b)}_{\theta^{(j)}}[\mathbf A]\notag\\
=&2\pi\delta(\theta^{(i)}-\theta^{(j)}) \int_{0\leq W[\mathbf A]< 1}{\cal D}\mathbf A
{\rm e}^{-{\rm i}(\theta^{(i)}-\theta^{(j)}) W[\mathbf A]}\psi^{(a)*}_{\theta^{(i)}}[\mathbf A] \psi^{(b)}_{\theta^{(j)}}[\mathbf A]\notag\\
=&2\pi\delta(\theta^{(i)}-\theta^{(j)}) \delta_{ab}\,,
\label{prod:nonnormalizable}
\end{align}
where we have assumed the normalization
\begin{align}
\int_{0\leq W[\mathbf A]< 1}{\cal D}\mathbf A
\psi^{(a)*}_{\theta^{(i)}}[\mathbf A] \psi^{(a)}_{\theta^{(j)}}[\mathbf A]=1\,.
\end{align}

As advertised above, the properties given by Eqs.~\eqref{Psi:EV:U4}, \eqref{prod:nonnormalizable} are analogous to those of the so-called $\theta$-vacua that are commonly discussed in the literature,
\begin{align}\label{eq:usual_theta_vacua}
\left|\theta\right\rangle=\sum_{n\in\mathbb{Z}}\,{\rm e}^{i n\theta}|n\rangle\,.
\end{align}
In the equation above, the $|n\rangle$ are eigenstates of the operator corresponding to $W[A]$ with integer eigenvalues, and hence in direct relation to {classical states of minimal energy} given by pure gauge configurations $\mathbf{A}={\rm i} U^{-1}\nabla U$, for which Eq.~\eqref{CS:diff} implies that $W[\mathbf{A}]=\nu(U;S)\in\mathbb{Z}$. This means that the wave-functionals $\Psi_{|\theta\rangle}[\mathbf A]=\langle \bf A|\theta\rangle$ corresponding to the states of Eq.~\eqref{eq:usual_theta_vacua} only have support in classical minima of the energy functional. In ordinary quantum mechanics, it is well known that in the case of multiple classical vacua the wave functions of the stationary states have support away from the classical minima, so that the states \eqref{eq:usual_theta_vacua} are not expected to correspond to physical stationary states {and therefore are in fact not ``vacua''}. Despite the analogies with the states~\eqref{eq:usual_theta_vacua}, our present discussion is more general, as we never assume that the wave functionals $\Psi_{{\theta^{(i)}}}(\mathbf A)$ have support in classical field configurations of {minimal energy}  only. 

Going back to the general wave functionals  $\Psi_{{\theta^{(i)}}}(\mathbf A)$, a particular gauge fixing may be imposed such that the condition $0\leq W[\mathbf A]< 1$ is obeyed along with additional constraints. But then with the corresponding gauge fixed inner product as in Eq.~(\ref{inner:product:g.f.}), there is no sum over $\nu$ so that states with $\theta^{(i)}\not=\theta^{(j)}$ are no longer orthogonal.

At this point, one may see two apparent ways of resolving this conflict.
\begin{enumerate}[label=(\alph*)]
\item
\label{option:nonnormalizable}
We dismiss the requirement that there should be a gauge fixed inner product as in Eq.~(\ref{inner:product:g.f.}). Instead, we take~(\ref{prod:nonnormalizable}) as an inner product even though we integrate over gauge-redundant configurations, which as mentioned before leads to  states that are not properly normalizable. They are therefore not members of a Hilbert space, and taking these as physical states would be in conflict with the postulates of quantum mechanics.
\item
\label{option:normalizable}
We stick with the inner product~(\ref{inner:product:g.f.}). To avoid a contradiction, only states $\Psi_{\theta^{(i)}}$ with a particular value of $\theta^{(i)}$ (modulo $2\pi$) can be in the Hilbert space. We will show that only one value of $\theta^{(i)}$ (modulo $2\pi$) is admissible, namely the one that leads to the prediction of no $CP$ violation in the strong interactions.  
\end{enumerate}

The way we have put things here anticipates that in the present work, we take the view that Option~\ref{option:normalizable} is the correct path to pursue. Nonetheless, we have to take due account of the fact that the vast majority of papers (see e.g. Ref.~\cite{DiLuzio:2020wdo} for a review), save Refs.~\cite{Okubo:1992na,Gracia-Bondia:2022eor}, follow Option~\ref{option:nonnormalizable}, propose the use of inner products over gauge redundant configurations and are not concerned with nonnormalizable states that are considered as unphysical in different contexts. (At best, in quantum mechanics, momentum eigenstates in a translation-invariant background, which are not normalizable, constitute a basis of the physical Hilbert space. They can be used to construct normalizable wave packets which are the physical states. When taking a trace over momentum eigenstates, the ensuing factors $(2\pi)^3\delta^3(0)$ can be identified with the infinite volume $V$. It is not clear and may not be possible to interpret the improper normalization of the $\theta$-vacua in a corresponding way. In particular, wave-packet constructions, even if the wave packets are strongly peaked about a certain value of $\theta^{(i)}$, are in conflict with gauge invariance.) In addition to these objections, we therefore show explicitly that the restoration of gauge covariance after fixing the transition function requires that $\theta^{(i)}$ takes values in accordance with $CP$ conservation.

To make such an argument, we recall that, as stated above, any transition function $U_4$---also one that is not constant---on $\partial F_4$ may be imposed as a partial gauge fixing before quantization. In order to restore gauge covariance, we may therefore consider the family of states that is obtained by having all possible transition functions $U_4$ on $\partial F_4$, thus composing a Hilbert space with an extended basis. In this extended space, the Hamiltonian must act consistently, i.e. a simultaneous covariant gauge transformation of the Hamiltonian and its energy eigenstates should not change the eigenvalues. In other words, changing the transition function should not change the spectrum. This condition leads to a restriction of the possible values of $\theta^{(i)}$, which turns out to resolve the aforementioned issues with normalization and gauge fixing. A more direct way to obtain a gauge covariant construction with respect to the transition functions is of course to fix the latter only after quantization, which is the order in which we work in Sections~\ref{sec:Gauss} and~\ref{sec:first:quantize:then:constrain}.

Before proceeding, we comment on the case when space is $\mathbbm R^3$ and infinite as opposed to $T^3$ and finite. For infinite space, one may have reasons {to adopt the hypothesis that the gauge potentials must fall faster than $1/r$ at large distances}, which in turn limits the gauge transformations in Eq.~(\ref{gaugetrafo:spatial}) to those with $U(|\mathbf x|)\to \text{constant}$ for $|\mathbf x|\to \infty$~\cite{Jackiw:1979ur}. As a consequence, the points at infinity can then be identified so that space can be viewed as homeomorphic to the three-sphere $S^3$, for which $\nu(U,{S^3})\in\mathbb{Z}$. In particular, there are transformations $U^{(1)}$ with $\nu(U^{(1)},S^3)=1$ so that Eq.~(\ref{Psi:EV:U4}) holds for $U_4^{(1)}=U^{(1)}$. When we assume that gauge transformations that are not constant for $|\mathbf x|\to \infty$ are not permissible, we can also not discern which value of $\theta^{(i)}$ gives the correct basis states by studying the gauge-covariant behaviour of the states and the Hamiltonian under the restricted gauge transformations. But as we recall, for infinite spacetime, correlation functions computed in the functional approach turn out to be independent of $\theta$ and do not exhibit $CP$ violation~\cite{Ai:2020ptm}. (The {four-dimensional} volume of space is to be taken to infinity before summing over the topological sectors in the four-dimensional spacetime.) The choice of $\theta^{(i)}$ is then of no consequence for the predictions of observables. We can just fix it and therefore obtain a Hilbert space based on a gauge fixed inner product~(\ref{inner:product:g.f.}). {We shall return to the case of infinite $\mathbbm R^3$ in Section~\ref{sec:Minkowski}, subsequent to presenting the main results.}

Now, to further pursue Option~\ref{option:normalizable} we need to work out the covariant behaviour of the states under gauge transformations. Considering gauge transformations that change the transition function, this will fix the value of $\theta^{(i)}$ to be discussed in Section~\ref{sec:first:quantize:then:constrain}. Transformations that keep the transition function invariant are related with Gau{\ss}' law, as we recall next.

\section{Constraints on the quantum states}
\label{sec:quantum:constraints}

Gauge fixing as stated in Section~\ref{sec:transition:function} corresponds to the choice of a constrained hypersurface $\cal A$ for the field configurations $\mathbf A(\mathbf x)$ so that each gauge-inequivalent configuration appears one time and one time only. Under the induced inner product~(\ref{inner:product:g.f.}) {without restricting the gauge configurations to a certain space $\cal A$}, the space of square-integrable functionals $\Psi$, denoted here by $\mathcal{C}$, is generally no longer a Hilbert space. However, since in the present case the choice of $\cal A$ corresponds to the removal of gauge redundancies, we can expect that for given $\cal A$, there is a subspace $\subset \mathcal{C}$ that qualifies as a Hilbert space also for the inner product~(\ref{inner:product:g.f.}) and that is constituted by physical states.

To identify this subspace, we first notice that the independence of probabilities on the choice of $\cal A$ requires the following form of the wave-functional for a state labeled with $(a)$:
\begin{align}
\label{product:function:phase}
\Psi^{(a)}[\mathbf A]=\Psi_\text{g.i.}^{(a)}[\mathbf A]\exp({\rm i}\varphi^{(a)}[\mathbf A])\,,
\end{align}
where $\Psi_\text{g.i.}^{(a)}[\mathbf A]$ is gauge-independent and $\varphi^{(a)}[\mathbf A]$ is a real functional. That is, gauge transformations can induce at most a phase change in the wave functionals, as such phases cancel when computing physical transition or probability amplitudes defined from inner products of wave functions.

Furthermore, the Hamiltonian as well as the operator $-{\rm i}\delta/\delta \mathbf A$ that it is built from should remain Hermitian operators. This leads to 
\begin{align}
&\int_{\cal A}{\cal D}\mathbf A f_{\cal A}[\mathbf A]
\Psi^{(a)*}_{\text{g.i.}}[\mathbf A] \exp(-{\rm i}\varphi^{(a)}[\mathbf A])\frac{\delta}{{\rm i}\delta A(\sigma)}\Psi^{(b)}_{\text{g.i.}}[\mathbf A]\exp({\rm i}\varphi^{(b)}[\mathbf A])
\notag\\
=&\int_{\cal A}{\cal D}\mathbf A f_{\cal A}[\mathbf A]
\left(\frac{\delta}{{\rm i}\delta A(\sigma)} \Psi^{(a)}_{\text{g.i.}}[\mathbf A] \exp({\rm i}\varphi^{(a)}[\mathbf A])\right)^*\Psi^{(b)}_{\text{g.i.}}[\mathbf A]\exp({\rm i}\varphi^{(b)}[\mathbf A])\,.
\end{align}
Note that the above equality must hold for any pair of states such that $\Psi^{(a)}$ and $\Psi^{(b)}$ are not necessarily orthogonal here. Therefore, it must be that $\varphi^{(a)}[\mathbf{A}]=\varphi^{(b)}[\mathbf{A}]=\varphi[\mathbf{A}]$. For the terms with derivatives acting on $\varphi[\mathbf A]$, the above equation holds because $\varphi[\mathbf A]$ is independent of $(a)$. This explains the Hermiticity of all derivatives in the directions $\sigma=\sigma_{\text{gauge}}$ because the remaining factor $\Psi_\text{g.i.}^{(a)}[\mathbf A]$ is gauge invariant. For all other directions $\sigma$ the variational derivative is equivalent to one that is tangential to $\cal A$. By the translation-invariance of ${\cal D}[\mathbf A]f_{\cal A}[\mathbf A]$ on $\cal A$, the above relation then follows from partial integration and vanishing boundary terms involving $|\Psi^{(a)*}_{\text{g.i.}}[\mathbf A]|^2$. {In some directions in field space, there arise improper boundaries when the physical field strengths become unbounded and the wave functional goes to zero. In other particular directions boundaries occur for finite physical fields from e.g. the constraint $0\leq W[\mathbf A]<1$. The configurations on such gauge-equivalent locations on the hypersurfaces $W[\mathbf A]=0$ and $W[\mathbf A]=1$ are related by gauge transformations which leave $|\Psi^{(a)*}_{\text{g.i.}}[\mathbf A]|^2$ invariant, so that the contributions from the hypersurfaces cancel}.  We conclude
\begin{align}
\label{product:function:phase2}
\Psi^{(a)}[\mathbf A]=\Psi_\text{g.i.}^{(a)}[\mathbf A]\exp({\rm i}\varphi[\mathbf A])\,,
\end{align}
and note that $\varphi[\mathbf A]$ is the only piece that {still} may depend on the choice of ${\cal A}$.

We proceed to show that there are indeed solutions for the wave functional that factorize into a piece $\Psi_\text{g.i.}^{(a)}[\mathbf A]$ that is gauge independent and another piece $p[\mathbf A]$ that may be gauge dependent but is universal for all states, i.e.
\begin{align}
\label{product:function}
\Psi^{(a)}[\mathbf A]=\Psi_\text{g.i.}^{(a)}[\mathbf A]\,p[\mathbf A]\,,
\end{align}
which is a more general form of Eq.~(\ref{product:function:phase2}) and hence a necessary condition.
For this purpose, we use the fact that
\begin{align}
\label{eq:W:fact}
\frac{\delta}{\delta \mathbf A(\mathbf x)} W[\mathbf A]=\frac{g}{8\pi^2}\mathbf B(\mathbf x)\,,
\end{align}
in order to carry out the following change of basis:
\begin{align}
 \Psi^\prime[\mathbf A]=&{\rm e}^{-{\rm i}(\theta-\theta_\Pi) W[\mathbf A] }\Psi[\mathbf A]\,,
 \label{Psi:prime}
\\
{ H}^\prime=&{\rm e}^{-{\rm i}(\theta-\theta_\Pi) W[\mathbf A] }{H}{\rm e}^{{\rm i}(\theta-\theta_\Pi) W[\mathbf A] }=
\frac{1}{2}\,\int{\rm d}^3 x\,{\rm tr}\left[-g^2\frac{\delta^2}{\delta\mathbf A^2}+\mathbf B^2\right
]\notag\\
=&-\frac{g^2}{2}\int\limits_\sigma\hskip-.5cm\scalebox{1.}{$\sum$}  \frac{\delta^2}{\delta A^2(\sigma)}  +\frac{1}{2}\int{\rm d}^3 x\,{\rm tr}\,\mathbf B^2\label{H:prime}\,.
\end{align}
Now in the Hamiltonian $H^\prime$, as opposed to Eq.~(\ref{Hamiltonian:gauge}), there are no cross terms $\mathbf B(\mathbf x)\cdot (\delta/\delta\mathbf A(\mathbf x))$, as we make explicit within the last expression in Eq.~(\ref{H:prime}). {In what follows, we thus refer to the basis of the Hilbert space given by $\Psi'[\mathbf{A}]$ as the diagonal basis.} To separate field directions related to gauge transformations, it is useful to write
\begin{align}
 \int\limits_\sigma\hskip-.5cm\scalebox{1.}{$\sum$}  \frac{\delta^2}{\delta A^2(\sigma)}=\int\limits_{\sigma_{\rm gauge}}\hskip-.7cm\scalebox{1.}{$\sum$}  \frac{\delta^2}{\delta A^2(\sigma_{\rm gauge})}+\int\limits_{\sigma_{\parallel}}\hskip-.5cm\scalebox{1.}{$\sum$}  \frac{\delta^2}{\delta A^2(\sigma_{\parallel})}.
\end{align}
The directions parameterized by the $\delta A(\sigma_\parallel)$ correspond to field variations $G^a_i(\mathbf x;\sigma_\parallel)$ {(cf. Eq.~\eqref{eq:trafo:FD})} orthogonal to gauge transformations. Note that in general, $\delta A(\sigma_\parallel)$ is not tangent to the constrained surface $\cal A$ as such tangents may be linear combinations of $\delta A(\sigma_\parallel)$ and $\delta A(\sigma_{\rm gauge})$. However,  $\delta A(\sigma_\parallel)$ can be projected onto ${\cal A}$, upon which a gauge transformation is added. As $\Psi_{\rm g.i.}^{(a)}[\mathbf A]$ is gauge independent by definition, integrating the Schr\"{o}dinger equation in directions $\delta A(\sigma_\parallel)$ from a point ${\bf A}\in \A$ also amounts to finding the solution at $\mathbf A+\delta \mathbf A\in\mathcal A$ where $\delta {\bf A}$ is $\delta A(\sigma_\parallel)$ projected on $\A$. It should be kept in mind here that the separation into  $\delta A(\sigma_{\rm gauge})$ and $\delta A(\sigma_\parallel)$ depends on the point $\mathbf A\in {\cal A}$.

In the basis of Eq.~\eqref{Psi:prime}, the functional Schr\"odinger equation is manifestly separable and can be solved by a product ansatz. In
\begin{align}
\Psi'^{(a)}[\mathbf A]={\Psi'}_\text{g.i.}^{(a)}[\mathbf A]\,p'[\mathbf A]\,,
\end{align}
we take
\begin{align}
\label{product:ansatz:indpendences}
\frac{\delta}{\delta A(\sigma_\text{gauge})} \Psi_\text{g.i.}^{(a)}[\mathbf A]=0\,,\qquad
\frac{\delta}{\delta A(\sigma_\parallel)} p^{\prime (a)}[\mathbf A]=0
\end{align}
and first solve the eigenvalue problem for the factor $p^\prime[\mathbf A]$
\begin{align}
\label{eq:delta-p-prime}
    \int\limits_{\sigma_{\rm gauge}}\hskip-.7cm\scalebox{1.}{$\sum$}  \frac{\delta^2}{\delta A^2(\sigma_{\rm gauge})} p'[\mathbf A]=&\,\mu\, p'[\mathbf A]\,,
\end{align}
where $\mu$ is real due to Hermiticity. The first equality in Eq.~(\ref{product:ansatz:indpendences}) is in accordance with the general requirement~(\ref{product:function}) from gauge invariance. The Schr\"odinger equation  corresponding to a state with energy $E^{(a)}$ then gives
 \begin{align}
  \left(-\frac{g^2}{2}\int\limits_{\sigma_{\parallel}}\hskip-.5cm\scalebox{1.}{$\sum$}  \frac{\delta^2}{\delta A^2(\sigma_{\parallel})}
% %%%%
 +\frac{1}{2}\,\int {\rm d}^3x\, {\rm tr} \mathbf B^2\right){\Psi'}_\text{g.i.}^{(a)}
 %%%%
 =\left(E^{(a)}+\frac{g^2\mu}{2}\right){\Psi'}_\text{g.i.}^{(a)}\,,
\end{align}
where we have used Eqs.~(\ref{product:ansatz:indpendences}) and~(\ref{eq:delta-p-prime}). Note that since the transformation in Eq.~(\ref{Psi:prime}) is a phase, we have shown the existence of solutions of the form~(\ref{product:function:phase2}) not only for the diagonal but also for a general basis.

The solutions to Eq.~(\ref{eq:delta-p-prime}) however cannot be of the form $p'[\mathbf A]=\exp({\rm i}\varphi[\mathbf A])$ assumed in Eq.~(\ref{product:function:phase2}) unless $\varphi[\mathbf A]\equiv 0$. This is because the solutions only depend on the gauge transformation and they can be written as product functions with respect to the particular directions $\sigma_{\text{gauge}}$. If they are pure phases, the derivatives in the directions of particular generators of gauge transformations must be constant for solutions to Eq.~(\ref{eq:delta-p-prime}). Hence, they give rise to a representation of the gauge group
\begin{align}
\exp({\rm i}\varphi[\mathbf{A}_{U_3}])=\exp({\rm i}\varphi[\mathbf{A}_{U_2}]) \exp({\rm i}\varphi[\mathbf{A}_{U_1}])\,,\quad \text{where} \quad U_3=U_2 U_1\,,
\end{align}
so that in particular $\varphi[\mathbf{A}_{U_2 U_1}-\mathbf{A}_{U_1 U_2}]=0$.
For a non-Abelian gauge group, this is only possible if $\varphi[\mathbf{A}_U]\equiv 0$.

Note that this argument can also be used to prove that the eigenfunctions of the Laplacian on a sphere of two or more dimensions, i.e. the spherical harmonics, cannot be pure phases over the whole sphere except for the constant solution with eigenvalue zero.

In summary, in the diagonal basis~(\ref{H:prime}) for the Hamiltonian, the only permitted eigenvalue of
\begin{align}
\int\limits_{\sigma_\text{gauge}}\hskip-.7cm\scalebox{1.}{$\sum$} \frac{\delta^2}{\delta A(\sigma_{\rm gauge})^2}
\end{align}
under the condition that gauge transformations only induce pure phases on the wave functional is $\mu=0$, and  the phases must be constant. That is
\begin{align}
\label{inv:Psi:prime}
\frac{\delta}{\delta  A(\sigma_{\rm gauge})} \Psi^\prime[\mathbf A]=\int{\rm d}^3 y \sum\limits_{i,a} G^a_i(\mathbf{y}; \sigma_\text{gauge}) \frac{\delta}{\delta A^a_i(\mathbf y)}\Psi^\prime[\mathbf A]=0\,.
\end{align}
In the diagonal basis of Eq.~\eqref{Psi:prime}, this fixes the wave functional to be invariant under all gauge transformations that are continuously connected with the identity.

Note that in the Hilbert space of functionals with the property~(\ref{inv:Psi:prime}) it is possible to evaluate the action of $H^\prime$ (i.e. in the diagonal basis) entirely in terms of derivatives $\delta/\delta A(\sigma)$ pointing in directions that are  tangential on any chosen $\cal A$ without changing the spectrum of the Hamiltonian by this restriction. To isolate these directions in the individual terms that add up to the Hamiltonian, the absence of the cross term between the magnetic field and the functional derivative in Eq.~(\ref{H:prime}) has been necessary.

As a consequence of the above, we can state that only derivatives
\begin{align}\lim_{\varepsilon\to 0} \frac1\varepsilon\left(\Psi^\prime[\mathbf A+\varepsilon\Delta \mathbf A]-\Psi^\prime[\mathbf A]\right)\end{align}
for $\Delta\mathbf A$ that are tangential on $\cal A$ may contribute to observables such as the Hamiltonian. As follows from the  line of reasoning in this section, this principle is equivalent to maintaining Hermiticity of the corresponding operators in the space of physical states. We would therefore like to see whether the principle can also be phrased in terms of some notion of tangential derivatives in the case of  $H$ in the general basis. For this purpose, define
\begin{align}
\mathbf D_{\mathbf A} \Psi[\mathbf A]={\rm i}\left(\frac{\delta}{{\rm i}\delta \mathbf A}-(\theta-\theta_\Pi)\frac{g}{8\pi^2}\mathbf B\right) \Psi[\mathbf A]\,.
\end{align}
The operator $\mathbf D_{\mathbf A}$ is of the same form as the momentum operator~(\ref{can:mom:gauge}). It therefore satisfies the commutation relations~(\ref{eq:canonical:commutators}) with the field operator $\mathbf A$ which is the appropriate algebraic property to use it as a generator of translations in  $\mathbf A$. When we denote the associated translation operator as $T[\Delta \mathbf A]$, it acts on $\Psi[\mathbf A]$ as
\begin{align}
T[\Delta\mathbf A]\Psi[\mathbf A]=\exp\left\{-{\rm i}(\theta-\theta_\Pi)\left(W[\mathbf A+\Delta\mathbf A]-W[\mathbf A]\right)\right\}\Psi[\mathbf A+\Delta \mathbf A]
\end{align}
as can be seen when using Eq.~(\ref{eq:W:fact}) and considering infinitesimal translations, for which
\begin{align}
T[\varepsilon\Delta \mathbf A]\Psi[\mathbf A]\approx&
\exp\left(-{\rm i} (\theta-\theta_\Pi) \frac{g}{8\pi^2} \int{\rm d}^3y\,\varepsilon\Delta\mathbf A\cdot \mathbf B\right)\Psi[\mathbf A+\varepsilon\Delta \mathbf A]\notag\\
\approx&\Psi[\mathbf A]+\int{\rm d}^3 y\,\varepsilon\Delta \mathbf A  \cdot \mathbf D_{\mathbf A} \Psi[\mathbf A]\,,
\end{align}
or, alternatively,
\begin{align}
\int{\rm d}^3 y\,\Delta \mathbf A  \cdot \mathbf D_{\mathbf A} \Psi[\mathbf A]=\lim_{\varepsilon\to 0}\frac{T[\varepsilon\Delta \mathbf A]\Psi[\mathbf A]-\Psi[\mathbf A]}{\varepsilon}\,,
\end{align}
i.e. $\mathbf D_{\mathbf A}$ is indeed a generator of $T$. Now in terms of $\mathbf D_{\mathbf A}$ and $\Psi$, the condition~(\ref{inv:Psi:prime}) can be written as
\begin{align}
\label{inv:Psi}
\frac{\delta}{\delta  A(\sigma_{\rm gauge})}\Psi^\prime[\mathbf A]=\int{\rm d}^3 y\, {\rm e}^{-{\rm i}(\theta-\theta_\Pi)W[\mathbf{A}]}\sum\limits_{i,a} G^a_i(\sigma_\text{gauge},\mathbf y) [D_{\mathbf A}]^a_i\Psi[\mathbf A]=0,
\end{align}
while the Hamiltonian can be expressed as follows,
\begin{align}
H=\frac12\int{\rm d}^3 x \,{\rm tr}\left[-g^2{\mathbf D}_{\mathbf A}^2+\mathbf{B}^2\right]\,.
\end{align}
The condition of Eq.~\eqref{inv:Psi} can be interpreted in terms of generalized derivatives ${\rm D}/{\rm D}A(\sigma_{\rm gauge})$ in the directions of gauge transformations, with
\begin{align}
\frac{{\rm D}}{{\rm D}A(\sigma)}\equiv\int{\rm d}^3 y\, {\rm e}^{-{\rm i}(\theta-\theta_\Pi)W[\mathbf{A}]}\sum\limits_{i,a} G^a_i(\mathbf y;\sigma) [D_{\mathbf A}]^a_i.
\end{align}
In terms of the wave functionals $\Psi$, these may only have nonvanishing derivatives for ${\rm D}/{{\rm D}A(\sigma_\parallel)}$ associated with $G^a_i(\mathbf y;\sigma_\parallel)$ orthogonal to the infinitesimal gauge transformations $G^a_i(\mathbf y;\sigma_{\rm gauge})$.

Hence, provided the condition~(\ref{inv:Psi}) or, equivalently, Eq.~(\ref{inv:Psi:prime}), there are only contributions to the Hamiltonian $H$ from changes in the wave functional $\Psi[\mathbf A]$ under the translations $T[\Delta \mathbf A]$ for $\Delta \mathbf A$ that are tangent on any given $\cal A$. As long as this condition is satisfied for some translation operator $T$ and the Hilbert space is restricted correspondingly, the observables do not depend on the choice of $\cal A$ and gauge fixing following Eq.~(\ref{inner:product:g.f.}) is thus possible.

\section{Gau{\ss}' law constraint}
\label{sec:Gauss}

The presence of some of the unphysical states in the spectrum of the Hamiltonian is not a problem of quantization in the first place but can be traced to the fact that the canonical equations of motion for the Hamiltonian~(\ref{Hamiltonian:gauge}) do not contain Gau{\ss}' law. As it turns out, imposing Gau{\ss}' law, one restricts the Hilbert space in such a way that the unphysical states are removed and that it allows for gauge fixing. These matters are well known for gauge theory in Minkowski spacetime~\cite{Jackiw:1979ur}. Here, we shall review these points in view of what happens on the three-torus $T^3$. We add to the discussion the point of view that one can also argue conversely that gauge invariance implies that the physical states when constrained according to the criteria derived in Section~\ref{sec:quantum:constraints} satisfy Gau{\ss}' law.

In the temporal gauge, Hamilton's equations do not yield Gau{\ss}' law. It can be enforced at the level of the quantized theory by imposing
\begin{align}
\label{eq:Gauss:law}
\mathbf D\cdot\mathbf E^a\, \Psi[\mathbf A]=0
\end{align}
on the physical states represented by the wave functionals $\Psi[\mathbf A]$. In general, there will be states that do not observe this condition. Those will be removed from the Hilbert space of physical states.

For this construction to make sense, the condition~(\ref{eq:Gauss:law}) must define a subspace that is invariant under the time evolution. To show this is the case, one takes the gauge transformations~(\ref{gaugetrafo:spatial}) for symmetries with pertaining Noether charges. For an infinitesimal generator $\Omega(\mathbf x)$ of the transformation considered, the Noether charge is given by
\begin{align}
Q(\Omega)
=&\frac{1}{g} \int {\rm d}^3 x\,{\rm tr}\left[\Pi^i (D^i \Omega)\right]
= \int_V{\rm d}^3 x \,{\rm tr}\left[\left(-E^i+\frac{g^2}{8\pi^2}\theta B^i \right) D^i \Omega\right]\notag\\
=& \int{\rm d}^3 x\, {\rm tr}\left[\Omega D^i \left(E^i-\frac{g^2}{8\pi^2}\theta B^i \right)\right]
+\int_{\partial V} {\rm d} a^i\, {\rm tr}\left[\Omega\left(-E^i+\frac{g^2}{8\pi^2}\theta B^i \right) \right]\,.
\label{Q:Omega}
\end{align}
Here, we have integrated by parts, $V$ denotes the integration volume, $\partial V$ its boundary and $\mathbf a$ a normal surface element. In the present case, we take $V=F_4$, which can be identified with the cube that yields $T^3$ when its opposite faces are identified.

We shall assume for now that the surface term in Eq.~(\ref{Q:Omega}) vanishes. This is guaranteed if $\Omega(\mathbf x)=0$ for $\mathbf x\in\partial V$. Since $D^i E^i\,\Psi=0$ for
the states that observe Eq.~(\ref{eq:Gauss:law}) and $D^i B^i=0$ per Eq.~(\ref{B:field}), it follows that $Q(\Omega)\,\Psi=0$. Since $Q(\Omega)$ is the operator corresponding to the Noether charge associated with gauge transformations, it is a generator of the latter in the Hilbert space of wave functionals. So under gauge transformations that are connected with the identity and that are generated by some $\Omega$ that vanishes on $\partial V$, the $\Psi$ satisfying Eq.~(\ref{eq:Gauss:law}) are invariant. Since the $Q(\Omega)$ are Hermitian, states with a common eigenvalue of $Q$ define a subspace of the Hilbert space of wave functionals $\Psi[\mathbf A]$. And Furthermore, since these are Noether charges,
\begin{align}
\label{H:Q:commute}
[H,Q(\Omega)]=0
\end{align}
so that these subspaces have bases made up from eigenstates of the Hamiltonian and are thus invariant under time evolution. As $\Omega(\mathbf x)$ is arbitrary (up to the boundary condition $\Omega(\mathbf x)=0$ on $\partial V$), the states that satisfy Gau{\ss}' law~(\ref{eq:Gauss:law}) indeed constitute a subspace that is invariant under the Hamiltonian evolution. The requirement $\Omega({\bf x})=0$ on $\partial V$ is equivalent to that $U_4(\bf{x})$ is constant there.

We can also turn the argument around, i.e. instead of imposing Gau{\ss}' law to derive how the states behave under gauge transformations we can obtain Gau{\ss}' law by imposing gauge invariant observables as discussed in Sections~\ref{sec:gauge:fixing} and~\ref{sec:quantum:constraints}. To that end, we now consider gauge transformations in directions that we label with $\sigma_{\text{Gau{\ss}}}$, that have $\Omega=0$ on $\partial V$ and therefore give no boundary term in Eq.~(\ref{Q:Omega}). Note that we can relate this with $Q(\Omega)$ from Eq.~(\ref{Q:Omega}) as
\begin{align}
\label{dirder:periodic:Omega}
\frac{\delta}{\delta A(\sigma_{\text{Gau{\ss}}})}\Psi[\mathbf A]={\rm i}\,Q(\Omega)\Psi[\mathbf A]\,.
\end{align}

Equation~(\ref{inv:Psi:prime}) for $\sigma_{\text{gauge}}=\sigma_{\text{Gau{\ss}}}$ reads
\begin{align}
\frac{\delta}{\delta A(\sigma_{\text{Gau{\ss}}})}\Psi^\prime[\mathbf A]=0\,.
\end{align}

Next note that $W[\mathbf A]$ is invariant under gauge transformations $U$ connected with the identity that are constant on the surface $\partial F_4$. To see this, we apply the Gau{\ss} theorem to the surface term in Eq.~(\ref{CS:diff})) as
\begin{align}
-\frac{{\rm i}}{8\pi^2}\varepsilon_{ijk}\int_{S}{\rm d}^3x\,\partial_j{\rm tr}\left[A_k(\partial_i U) U^{-1}\right]=-\frac{{\rm i}}{8\pi^2}\int\limits_{\partial S}{\rm d}a_j \varepsilon_{ijk}{\rm tr}\left[A_k(\partial_i U) U^{-1}\right]=0\,.
\end{align}
for $S=F_4$. As both $\mathbf a$ and $\nabla U$ are normal vectors on $\partial S$, the scalar triple product is zero. So since
\begin{align}
\frac{\delta}{\delta A(\sigma_{\text{Gau{\ss}}})}W[\mathbf A]=0\,,
\label{inv:W}
\end{align}
it also follows that
\begin{align}
\frac{\delta}{\delta A(\sigma_{\text{Gau{\ss}}})}\Psi[\mathbf A]=0\,.
\end{align}
By Eq.~(\ref{dirder:periodic:Omega}), this also means $Q(\Omega)\Psi[\mathbf A]=0$. As $\Omega$ is an arbitrary function vanishing on $\partial V$, by Eq.~(\ref{Q:Omega}), Gau{\ss}' law~(\ref{eq:Gauss:law}) is implied.

\section{Gauge covariance with respect to general changes of the transition function}
\label{sec:first:quantize:then:constrain}

Now, we turn to generators $\Omega$ that do not vanish on $\partial V$. Taking for $V=F_4$,  the pertaining transformations in general change the transition function on $\partial F_4$. This implies that $U_4$ now can change in a general way, i.e. it is not fixed on $\partial F_4$. We label these transformations with $\sigma_{\partial F_4}$. The discussion is in large parts similar to the one for generators vanishing on $\partial V$ from Section~\ref{sec:Gauss} but there are also crucial differences.

First, we notice that we can still associate with $\Omega$ the Noether charges $Q(\Omega)$ in Eq.~(\ref{Q:Omega}). In the present case, in contrast to Section~\ref{sec:Gauss}, we can no longer assume that the surface term vanishes.

Once again, we can perceive $\Omega$ as a generator of a gauge transformation---there is no principle forcing us to have periodic gauge potentials even on the torus. Therefore, we can determine what gauge invariance tells us about the form of the wave functionals $\Psi[\mathbf A]$. Again, this can be done by imposing that one can consistently fix the gauge by choosing a manifold $\cal A$, that the operators corresponding to observables remain Hermitian under the inner product~(\ref{inner:product:g.f.}) and that the observables do not depend on the particular choice of $\cal A$. As worked out in Section~\ref{sec:quantum:constraints}, this requires that we can omit derivatives in the direction of transformations generated by $\Omega$ from the Hamiltonian, i.e. Eq.~(\ref{inv:Psi:prime}).

We first carry out the transformation given by Eqs.~(\ref{Psi:prime}) and~(\ref{H:prime}). Equation~(\ref{inv:Psi:prime}) with $\sigma_{\text{gauge}} \rightarrow \sigma_{\partial F_4}$ reads
\begin{align}
\label{inv:Psi:g.f.}
\frac{\delta}{\delta \mathbf A(\sigma_{\partial F_4})}\Psi^\prime[\mathbf A]=0\,.
\end{align}
One essential difference with the discussion in Section~\ref{sec:Gauss} is that this relation now entails information on the behaviour of the wave functional under translations of the field by the transition function $U^{(1)}_4$ in Eq.~(\ref{Psi:EV:U4}). Since the transition functions are not required to be constant on $\partial F_4$, the same holds true for the allowed gauge transformations. Crucially, with gauge transformations  unrestricted, and with general transformations in a simple gauge group being continuously connected with the identity, the usual large gauge transformations that are constant on $\partial F_4$ and change $W[\mathbf A]$ by an integer are therefore connected with the identity, which allows to relate these with the infinitesimal generators~$\Omega$.

Any transition function $U_4^{(1)}$ that changes $W[\mathbf A]$ by one unit can therefore be composed from a sequence of infinitesimal transformations $\mathbbm 1+{\rm i} \Omega_K$, $K=1,2,\ldots$. The $\Omega_K$ generally do not vanish on $\partial V=\partial F_4$ even though $U_4^{(1)}$ does. In fact, it is essential that some $\Omega_K$ do not vanish on the boundary because otherwise they could not compose a transition function that changes $W[\mathbf A]$. But then the invariance of Eq.~(\ref{inv:Psi:g.f.}) of the wave functional under infinitesimal transitions implies that it is also invariant under large gauge transformations,
\begin{align}
\label{g.t.:Psi:prime}
\Psi^\prime[\mathbf A_{(U_4^{(1)})^n}]=\Psi^\prime[\mathbf A]\quad\text{where}\quad n\in\mathbbm Z\,.
\end{align}
From the above, we conclude that in the primed basis the wave functional is gauge invariant,
\begin{align}\label{eq:Psi_prime_gi}
 \Psi'[{\bf A}]=\Psi'_{\rm g.i.}[{\bf A}]\,.
\end{align}
Another essential difference with the case of transformations that vanish on $\partial V=\partial F_4$  is now that there is no relation corresponding to Eq.~(\ref{inv:W}) when taking $\sigma_{\partial F_4}$ instead of $\sigma_{\text{Gau{\ss}}}$. Hence, the invariance~\eqref{inv:Psi:g.f.} does not assure that $\delta \Psi({\bf A})/\delta {\bf A}(\sigma_{\partial F_4})=0$. In fact, since $U_4^{(1)}$ adds one unit to $W[\mathbf A]$, from Eq.~\eqref{Psi:prime} we can conclude that
\begin{align}
\label{eq:conclusion}
\Psi[\mathbf A_{(U_4^{(1)})^n}]={\rm e}^{{\rm i}n(\theta-\theta_\Pi)}\Psi[\mathbf A]\,.
\end{align}
Hence, in the original basis the phase acquired by the states under large gauge transformations is tied to the $CP$-odd parameters in the Hamiltonian.
Using Eqs.~(\ref{Psi:prime}) and~(\ref{H:prime}), this phase can be simultaneously removed from the Hamiltonian and the wave functional so that there remains no $CP$-odd parameter.

We have thus seen how gauge invariant observables determine the eigenvalue in Eq.~(\ref{Psi:EV:U4}). This leads to a restriction of the Hilbert space that now consists of normalizable states under the inner product~(\ref{inner:product:g.f.}), in contrast to Eq.~(\ref{prod:nonnormalizable}). To demonstrate this necessary restriction of the Hilbert space we have used the fact that the transition functions (and hence the gauge potentials) are not required to be constant on $\partial F_4$.

Equation~\eqref{eq:conclusion} does not depend on whether the Euclidean time is compact or not. Therefore, it applies for the spectrum of QCD on a three-torus at both zero and finite temperatures. Of course, at zero temperature, one could deduce the $CP$ conservation from the perspective given in Refs.~\cite{Ai:2020ptm,Ai:2023X}.

It has been noted in Ref.~\cite{Tokarev:1993eh} that gauge invariance of the wave function corresponding to Eq.~(\ref{eq:Psi_prime_gi}) can explain the absence of parity violation. However, the gauge transformations in that work are postulated and not derived.

Since the Hamiltonian $H'$ as in Eq.~\eqref{H:prime} is equivalent to an ordinary Hamiltonian without a $\theta$-term, we can therefore conclude that the partition function corresponding to finite temperature field theory is given by
\begin{align}\label{eq:Ztorusfinal}
Z=\sum\limits_{\Delta n=-\infty}^{\infty}\int \mathcal{D} A_{\Delta n}
{\rm e}^{-\frac{1}{2 g^2}\int_{T^4} {\rm d^4} x {\rm tr}F_{\mu\nu}F_{\mu\nu}{-S_{\rm E, g.f.}-S_{\rm E, ghost}}}\,,
\end{align}
where $S_{\rm E, g.f.}, S_{\rm E,ghost}$  are the gauge-fixing and ghost contributions and an additional integral over the ghost fields should be understood. The derivation of the partition function from the canonically quantized theory is outlined in Appendix~\ref{Appendix:path:integral}.  Here, the phase $\theta$ from the Lagrangian~(\ref{QCDlagrangian}) does not appear, having been cancelled by the correlated phases of the wave functionals of the states in the physical Hilbert space, leading to a partition function that is manifestly $CP$ conserving. This is in contrast to the partition functions usually implemented in lattice studies, such as in Refs.~\cite{Guo:2015tla, Shindler:2015aqa, Dragos:2019oxn, Giusti:2018cmp, Albandea:2024fui}, where $CP$-violation is sought after or found by calculating the path integral in finite volume and weighting the individual topological sectors with $\exp(-{\rm i}\theta)$. According to the present analysis, such setups do not correspond to an evaluation of the trace of the canonical density matrix over the states in canonically quantized QCD.

\section{Generalization to Minkowski spacetime}
\label{sec:Minkowski}

Throughout the previous sections of this article we have concentrated on the case of the torus in order to illustrate how the conclusion about $CP$ conservation in QCD is not necessarily restricted to an infinite spacetime or zero temperature. In contrast to this, in Ref.~\cite{Ai:2020ptm}, the lack of $CP$ violation in QCD in Minkowski spacetime at zero temperature is derived from a path integral formalism, in which the infinite spacetime limit is tied to the necessity of guaranteeing that the partition function of QCD is evaluated at the correct ground state, which requires sending the time interval to infinity.

Estimating the dependence of the vacuum partition function $\langle 0 | {\rm e}^{-{\rm i}HT}|0\rangle$ on the $CP$-odd parameters without taking $T\rightarrow\infty$ requires knowledge of the phase of the vacuum wave functional \cite{Ai:2022htq}, which becomes possible with the results of this work, enabling an independent derivation of the results of Ref.~\cite{Ai:2020ptm}.

The canonical quantization in Minkwoski spacetime goes along the same lines as on the torus. As before, one can define a physical Hilbert space by restricting the inner product to a gauge-fixed surface $\cal A$. By demanding independence of the observables with respect to the choice of $\cal A$, one concludes that the wave functionals are as in Eq.~\eqref{Psi:prime},
\begin{align}\label{eq:Psi_vs_Psiprime}
\Psi[\mathbf A]=&{\rm e}^{{\rm i}(\theta-\theta_\Pi) W[\mathbf A] }\Psi'[\mathbf A]\,,
\end{align}
with $\Psi'[\mathbf A]$ being invariant under gauge transformations that are continuously connected to the identity,
\begin{align}\label{eq:constraint_psi_prime}
 \frac{\delta}{\delta A(\sigma_{\rm gauge})}\Psi'[\mathbf A]=0\,.
\end{align}
The crucial observation is that, since gauge transformations are a redundancy, one should ensure invariance of  physical observables with respect to arbitrary gauge transformations, without restrictions. This is in contrast to common assumptions~\cite{Jackiw:1976pf,Jackiw:1979ur,Shifman:2022shi} which restrict gauge transformations  in the gauge $A^0=0$ to those satisfying 
\begin{align}\label{eq:U_constraint}
 U(|\mathbf{x}|)\rightarrow {\rm constant}\quad\text{for}\quad |\mathbf{x}|\rightarrow\infty\,.
\end{align}
Under the latter restriction, the functional $W[\bf A]$ in Minkowski spacetime changes by integer values under  large gauge transformations $(U^{(1)})^n$ that cannot be connected with the identity, which would then not fall under the constraint of Eq.~\eqref{eq:constraint_psi_prime}, thus allowing for $CP$-odd phases associated with the transformation properties of $\Psi'$ under the action of $U^{(1)}$.
As mentioned before, Eq.~\eqref{eq:U_constraint} is only a hypothesis that is discussed e.g. in the classic review~\cite{Jackiw:1979ur}. Motivated by the analysis for the torus case, we therefore lift the restriction of Eq.~\eqref{eq:U_constraint}. Once this is done, all gauge transformations can be connected to the identity, and demanding Eq.~\eqref{eq:constraint_psi_prime} leads to the same conclusion as for the torus, namely
\begin{align}
 \Psi'[\mathbf A]= \Psi'_{\rm g.i.}[\mathbf A]\Rightarrow \Psi[\mathbf A]={\rm e}^{{\rm i}(\theta-\theta_\Pi) W[\mathbf A] }\Psi'_{\rm g.i.}[\mathbf A]\,.
\end{align}

Again, there is no $CP$ violation, as is apparent when working in the primed basis in which the states are invariant under gauge transformations and the Hamiltonian does not depend on $\theta$. When working in a general basis, the $\theta$ dependence of the Hamiltonian is exactly correlated with the phases of the physical states, leading to cancellations in observables.

To make contact with the path integral results of Ref.~\cite{Ai:2020ptm}, we can proceed as in the case of the  torus and estimate the vacuum partition function at zero temperature as
\begin{align}\label{eq:Z_Minkowski}\begin{aligned}
 \langle 0| {\rm e}^{-{\rm i} HT}|0\rangle =\int {\cal D}{\bf A}_1({\bf x})\int {\cal D}{\bf A}_2({\bf x})
\Psi^{(0)*}[{\bf A}_2]\langle {\bf A}_2({\bf x})|{\rm e}^{-{\rm i} H{ T}}|{\bf A}_1({\bf x})\rangle\Psi^{(0)}[{\bf A}_1]\\
%%%%%
=\int {\cal D}{\bf A}_1({\bf x})\int {\cal D}{\bf A}_2({\bf x})
{\Psi}^{\prime(0)*}_{\rm g.i.}[{\bf A}_2]\langle {\bf A}_2({\bf x})|{\rm e}^{-{\rm i} H'{T}}|{\bf A}_1({\bf x})\rangle\Psi^{(0)\prime}_{\rm g.i.}[{\bf A}_1]\,,
\end{aligned}\end{align}
where the superscript $(0)$ indicates the ground state.
The result is a path integral with $\theta$ omitted from the Lagrangian, and with boundary conditions weighted by gauge-invariant vacuum wave functionals, which are $\theta$-independent as they solve Schr\"odinger equations with the Hamiltonian $H'$. In Ref.~\cite{Ai:2020ptm}, the vacuum partition function was estimated without the need to know the wave functionals $\Psi_{0}[\bf A]$ by leaving the boundary conditions free and  setting $T\rightarrow\infty$, which eliminates the contamination of excited states for an amplitude of the form $ \langle {\bf A}_2| \e^{-{\rm i} HT}|{\bf A}_1\rangle$ with arbitrary ${\bf A}_{1,2}$. This procedure should give a result equivalent to Eq.~\eqref{eq:Z_Minkowski} up to a normalization factor, and  indeed the $\theta$-dependence of the resulting partition function is shown to be an unphysical global factor which drops out of observables. Hence, the results of this article show that the limiting procedure of Ref.~\cite{Ai:2020ptm}, which is justified purely from the point of view of the path integral, is indeed projecting into the correct physical vacuum state, whose wave functional has a phase correlated with the $\theta$-term in the Hamiltonian, leading to a cancellation of the $\theta$-dependence in physical observables.

\section{The hidden parity symmetry of QCD in the presence of a $\theta$-term}
\label{sec:Parity}

Given the previous results about $CP$ conservation in the strong interactions, it is natural to wonder whether there is an explanation in terms of a symmetry of the theory, which differs from the usual parity transformation
\begin{align}\label{eq:P}
P:\quad {\bf x}\rightarrow -{\bf x}, \quad {\bf A}[{\bf x}]\rightarrow {\bf A}^P({\bf x})= -{\bf A}[-{\bf x}].
\end{align}

It turns out that such symmetry does exist. The Hamiltonian of Eq.~\eqref{Hamiltonian:gauge} commutes with a modified parity operator $\hat P$ with $\hat P^2$ {being} equal to a phase. {From this symmetry, we can deduce that energy eigenstates of the form~(\ref{eq:conclusion}), upon which the gauge-fixed Hamiltonian acts as a Hermitian operator, can be found to be also eigenstates of the usual parity $P$.} To construct $\hat P$, we start by considering large transformations $ U^{(1)}$ which change $W[\bf A]$ by one unit, either on the {three-torus or on  a hypersurface of constant time in Minkowski spacetime},
\begin{align}
 W[U^{(1)}{\bf A}]=W[{\bf A}]+1\,.
\end{align}
Under the parity transformations \eqref{eq:P}, one has
\begin{align}\label{eq:W_AP}
W[{\bf A}^P]=-W[{\bf A}]\,,
\end{align} as follows from Eq.~\eqref{CS:functional}. Then one has
\begin{align}\label{eq:WP}
\begin{aligned}
W[U^{(1)}(U^{(1)}{\bf A}^P)^P]=&\,W[(U^{(1)}{\bf A}^P)^P]+1=-W[ U^{(1)}{\bf A}^P]+1\\
%%%
=&\,-(W[{\bf A}^P]+1)+1=W[\bf A]\,.
\end{aligned}\end{align}
The functional $W[\bf A]$ is invariant under gauge transformations $U^{(0)}$ that are continuously connected with the identity, and {that go to $\mathbbm 1$ at} the boundary. By ``boundary'' here we refer either  to {$\partial F_4$} on the torus, or  the region $|\mathbf{x}|\rightarrow\infty$ {on a spatial hypersurface} in Minkowski spacetime. With $W[{\bf A}]=W[ U^{(1)}(U^{(1)}{\bf A}^P)^P]$  as in Eq.~\eqref{eq:WP}, it follows that for a generic $ U^{(1)}$, $\bf A$ and $ U^{(1)}(U^{(1)}{\bf A}^P)^P$ are related to each other by a gauge transformation connected with the identity. As in the class of large gauge transformations $U^{(1)}$ there is a freedom to multiply by transformations $U^{(0)}$, we can always find a representative $\widetilde U^{(1)}$ such that
\begin{align}\label{eq:Utilde}
   \widetilde U^{(1)}(\widetilde U^{(1)}{\bf A}^P)^P={\bf A}\,.
\end{align}
With this definition, one can consider  the following operator on wave functionals,
\begin{align}\label{eq:Pdef}
 \hat P\, \Psi[{\bf A}]= {\rm e}^{2{\rm i}(\theta-\theta_\Pi)W[{\bf A}]}\Psi[\widetilde U^{(1)}{\bf A}^P]\,.
\end{align}
The transformation $\hat P$ can be seen to be physically equivalent to an ordinary parity transformation, where the parity transformed gauge field ${\bf A}^P$ is gauge transformed by $\widetilde U^{(1)}$, and the wave functional acquires an extra phase. Both the additional gauge transformation and phase should not affect physical observables. The relevance of $\hat P$ is that it commutes with the Hamiltonian. Indeed, starting from Eqs.~\eqref{Hamiltonian:gauge} and~\eqref{eq:Pdef} one can write
\begin{align}\begin{aligned}
 &H(\hat P\,\Psi[\bf A])\\
 %%%%
 =&\int {\rm d}^3x\,\frac{1}{2}\left[\left(g\frac{\delta}{{\rm i}\delta \mathbf A^a}-\frac{g^2}{8\pi^2} (\theta-\theta_\Pi)\mathbf B^a\right)^2+(\mathbf B^a)^2\right]{\rm e}^{2{\rm i}(\theta-\theta_\Pi)W[{\bf A}]}\Psi[\widetilde U^{(1)}{\bf A}^P]\\
 %%%%
 =&\int {\rm d}^3x\, {\rm e}^{2{\rm i}(\theta-\theta_\Pi)W[{\bf A}]}\frac{1}{2}\left[\left(g\frac{\delta}{{\rm i}\delta \mathbf A^a}+\frac{g^2}{8\pi^2} (\theta-\theta_\Pi)\mathbf B^a\right)^2+(\mathbf B^a)^2\right]\Psi[\widetilde U^{(1)}{\bf A}^P]\\
 %%%%
 =&\int {\rm d}^3x\,{\rm e}^{2{\rm i}(\theta-\theta_\Pi)W[{\bf A}]}{\cal H[{\bf A}^P]}\Psi[\widetilde U^{(1)}{\bf A}^P]={\rm e}^{2{\rm i}(\theta-\theta_\Pi)W[{\bf A}]}( H\Psi)[\widetilde U^{(1)}{\bf A}^P]\\
 %%%%
 =&\hat P ({ H}\Psi[{\bf A}])\,.
\end{aligned}\end{align}
In the equation above, when going from the second to the third line we made use of Eq.~\eqref{eq:W:fact}, while in the third line we used parity invariance of the magnetic field, and gauge invariance of the Hamiltonian density, ${\cal H}[{\bf A}]={\cal H}[\widetilde U^{(1)}{\bf A}].$

The final property of $\hat P$ to be stated here is  that $\hat P^2$ evaluates to a constant phase, again reflecting that under $\hat P^2$ one should get a state which is physically equivalent to the original one. To see this, we use successive applications of the definition of Eq.~\eqref{eq:Pdef}, leading to
\begin{align}\begin{aligned}
 \hat P^2 \Psi[{\bf A}]=&\,\hat P ({\rm e}^{2{\rm i}(\theta-\theta_\Pi)W[{\bf A}]}\Psi[\widetilde U^{(1)}{\bf A}^P])\\
 %%%%
 =&\,{\rm e}^{2{\rm i}(\theta-\theta_\Pi)W[{\bf A}]}{\rm e}^{2{\rm i}(\theta-\theta_\Pi)W[\widetilde U^{(1)}{\bf A}^P]}\Psi[\widetilde U^{(1)}(\widetilde U^{(1)}{\bf A}^P)^P]\,.
\end{aligned}\end{align}
With $W[\widetilde U^{(1)}{\bf A}^P]=1+W[{\bf A}^P]=1-W[{\bf A}]$, and using Eq.~\eqref{eq:Utilde}, one gets
\begin{align}\label{eq:P2diag}
 \hat P^2 \Psi[{\bf A}]={\rm e}^{2\ii (\theta-\theta_\Pi)}\Psi[{\bf A}]\,.
\end{align}
A consequence of $[\hat P, H]=0$ is that the $\hat P,H$ can be diagonalized simultaneously. Hence, an eigenstate of $H$ can always be written as a sum of eigenstates of $\hat{P}$.  Writing an eigenstate of $H$ and $\hat P$ as in Eq.~\eqref{eq:Psi_vs_Psiprime}, it turns out that
\begin{align}\label{eq:hatPPsiprime}\begin{aligned}
 \hat P \Psi[{\bf A}]=&\,\hat P {\rm e}^{{\rm i}(\theta-\theta_\Pi) W[\mathbf A] }\Psi'[\mathbf A] 
 %%%
  ={\rm e}^{2{\rm i}(\theta-\theta_\Pi)W[{\bf A}]} {\rm e}^{{\rm i}(\theta-\theta_\Pi)W[\widetilde U^{(1)}{\mathbf A}^P] }\Psi'[\widetilde U^{(1)}{\mathbf A}^P]\\
  %%%
  =&\,{\rm e}^{{\rm i}(\theta-\theta_\Pi)}{\rm e}^{{\rm i}(\theta-\theta_\Pi) W[\mathbf A] }\Psi'[\widetilde U^{(1)}{\mathbf A}^P]\,.
\end{aligned}\end{align}
Given Eq.~\eqref{eq:P2diag}, the allowed eigenvalues of $\hat{P}$ are $\pm {\rm e}^{{\rm i}(\theta-\theta_\Pi)}$. In terms of the $\Psi'$, Eq.~\eqref{eq:hatPPsiprime}  means that for  eigenstates of $H$ and $\hat{P}$ one has
\begin{align}\label{eq:Psi'_P}
\Psi'[\widetilde U^{(1)}{\mathbf A}^P]=\pm \Psi'[{\mathbf A}] \Leftrightarrow \Psi'[{\mathbf A}^P]=\pm \Psi'[(\widetilde U^{(1)P})^{-1}{\bf A}]\,,
\end{align}
with  $U^{(1)P}({\bf x})\equiv U^{(1)}({-\bf x})$.  Using Eqs.~\eqref{CS:diff} and~\eqref{eq:nu}, one can see that
\begin{align}\label{eq:WtildeUinvA}
 W[(\widetilde U^{(1)P})^{-1}{\bf A}]=W[{\bf A}]+1\,.
\end{align}
Combining Eq.~\eqref{eq:Psi'_P} with Eq.~\eqref{eq:Psi_vs_Psiprime} and using Eqs.~\eqref{eq:W_AP} and~\eqref{eq:WtildeUinvA}  leads to the following identity for general eigenstates of the Hamiltonian and the $\hat P$ operator:
\begin{align}\label{eq:Parity_general}
 \Psi[{\bf A}^P]=\pm{\rm e}^{-{\rm i}(\theta-\theta_\Pi)(1+2W[{\bf A}])}\Psi[(\widetilde U^{(1)P})^{-1}{\bf A}]\quad\text{for general eigenstates of $H,\hat P$.}
\end{align}
Equation~\eqref{eq:Parity_general} has a profound implication: the $\hat P$ symmetry of the Hamiltonian implies that the effect of a parity transformation is equivalent to a rephasing of the wave functional, together with a large gauge transformation.   {Parity conservation can then be seen explicitly when additionally} imposing the constraints on the Hilbert space of Section~\ref{sec:quantum:constraints} that remove the gauge redundancy and ensure independence of the gauge fixing. As we have seen, the constraints  amount to $\Psi'$ being invariant under all gauge transformations, in particular under $\widetilde U^{(1)}$. Then Eqs.~\eqref{eq:Psi'_P} and \eqref{eq:Parity_general} lead to
\begin{align}\label{eq:parity_symmetry_physical}\begin{aligned}
 &\Psi'[{\mathbf A}^P]=\pm \Psi'[{\mathbf A}] \quad\text{for physical states, { diagonal} basis},\\
 %%%%
 &\Psi[{\mathbf A}^P]=\pm {\rm e}^{-2{\rm i}(\theta-\theta_\Pi)W[{\bf A}]}\Psi[{\bf A}] \quad\text{for physical states, general basis.}
\end{aligned}\end{align}
That is, in both the diagonal and the general basis, up to a phase the states are eigenvectors of parity, which thus remains  conserved under time evolution. In the {{primed} diagonal} basis one recovers the usual $\pm 1$ eigenvalues of $P$, while in the  general basis, the states  acquire an unobservable, $\theta$-dependent phase.

To conclude this section, let us discuss how this parity symmetry can be seen from the path integral formulation {once the projection to physical states has been correctly taken into account}. { To that end, we recall} that a proper path integration should weigh the boundary conditions with the vacuum wave functionals, as in Eqs.~\eqref{eq:Z_torus} and \eqref{eq:Z_Minkowski}. While a parity transformation of ${\bf A}\rightarrow {\bf A}^P$ is not a symmetry of the Lagrangian, so that the integrand $\exp({\rm i}S)$ acquires a phase, this is compensated by the phases acquired by the wave functionals under parity, as in  Eq.~\eqref{eq:parity_symmetry_physical}. Consider for example the partition function of \eqref{eq:Z_Minkowski}, where we replace the transition amplitude with an appropriate gauge-fixed path integration with boundary conditions fixed by ${\bf A}_1$ and ${\bf A}_2$,
\begin{align}\label{eq:Z_Minkowski_2}\begin{aligned}
 \langle 0| \e^{-{\rm i} HT}|0\rangle =\,\int {\cal D}{\bf A}_1({\bf x})\int {\cal D}{\bf A}_2({\bf x})\int_{{\bf A}_1}^{{\bf A}_2} {\cal D} {\bf A}(t,{\bf x})
{{\Psi}^{(0)*}[{\bf A}_2]\Psi^{(0)}[{\bf A}_1]}{\rm e}^{{\rm i } (S[{\bf A}]{+S_{\rm g.f.}+S_{\rm ghost}})}\,.
\end{aligned}\end{align}
The $\theta$-term in the action $S$ has to be correlated with the $CP$-odd quantity $\theta-\theta_\Pi$ appearing in the Hamiltonian  of Eq.~\eqref{Hamiltonian:gauge}. 
In the following, we choose $\theta_\Pi=0$, which leads to the usual path integral with
\begin{align}
 S\supset S_\theta = \frac{\theta}{16\pi^2}{\rm tr}\,\int {\rm d}^4x F_{\mu\nu}\tilde F^{\mu\nu}\,.
\end{align}
Choosing a gauge $A^0=0$ and with ${\bf A}\rightarrow 0$ for $|{\bf x}|\rightarrow\infty$, $S_\theta$ becomes
\begin{align}
 S_\theta[{\bf A}]=\theta \Delta n = \theta (W[{\bf A}_2]-W[{\bf A}_1])\,.
\end{align}
Under parity, $W[\bf A]$ changes sign, so that
\begin{align}
 {\rm e}^{{\rm i}S[{\bf A}^P]}={\rm e}^{-2{\rm i}\theta(W[{\bf A}_2]-W[{\bf A}_1)]} {\rm e}^{{\rm i}S[{\bf A}^P]}\,.
\end{align}
The extra phase in $ {\rm e}^{{\rm i}S[{\bf A}^P]}$ is exactly cancelled by the rephasings of the wave functionals in Eq.~\eqref{eq:Z_Minkowski_2} under parity transformations, as follows from Eq.~\eqref{eq:parity_symmetry_physical} with $\theta_\Pi=0$. Hence the partition function is invariant under parity, regardless of $\theta$. When one estimates the partition function without knowledge of $\Psi^{(0)}[{\bf A}]$ by sending  $T\rightarrow \infty$ and keeping boundary conditions free, as in Ref.~\cite{Ai:2020ptm}, the partition function should be invariant under parity up to a change of the unphysical normalization factor, which is indeed the case for the results of Ref.~\cite{Ai:2020ptm}. Again, the limiting procedure of that reference can be justified by requiring consistency with the results of the canonical formalism.

\section{Conclusions}
\label{sec:conclusions}

On $T^3$, The existence of eigenstates of the Hamiltonian of the Bloch form~(\ref{Psi:EV:U4}) relies on fixing the gauge so that the transition function $U_4=\mathbbm 1$ on $\partial F_4=\partial W_3$. This then implies that the vector potentials $\mathbf A$ are periodic on $T^3$. As this is not a general gauge, we have to check that the conclusions drawn from fixing the gauge in this way are gauge invariant {and correspond to expectation values of Hermitian operators}. These considerations reveal that the phase $\theta^{(i)}$ {associated with the periodicity of the Bloch state} is pinned in such a way that it cancels the phase $\theta-\theta_\Pi$ in the Hamiltonian. {Furthermore}, we can use an inner product that does not extend over gauge redundant configurations. As a consequence, the finite temperature field theory for the Lagrangian~(\ref{QCDlagrangian}) does not yield $CP$-odd correlation functions.

This constraint on $\theta^{(i)}$ also determines the Hilbert space of physical wave functionals. Within this space, the absence of $CP$ violation can be understood in terms of an exact symmetry of the QCD Hamiltonian which exists for arbitrary values of $\theta$ and can be viewed as a deformation of the ordinary parity symmetry by additional unphysical phases and gauge transformations.

We have focused here on the three-torus, in order to avoid a discussion of boundary conditions that one has to carry out at spatial infinity in Minkowski spacetime. If one does not adapt the common assumption that the relevant pure gauge configurations at infinity must satisfy $U(|\mathbf{x}|)\rightarrow {\rm constant}$ for $|\mathbf{x}|\rightarrow\infty$, the reasoning for the torus applies also to Minkowski spacetime, which is complementary to the findings for path integrals in infinite volume~\cite{Ai:2020ptm}.

In either case, Minkowski spacetime or three-torus, and in contrast to the usual $\theta$-vacua, the Hilbert space of physical states is by construction restricted in such a way that these are properly normalizable. It thus turns out that the topological configurations in QCD pose no obstacle to a  
straightforward quantum-mechanical interpretation of squared amplitudes as probabilities.

\section*{Acknowledgments}
The work of WYA is supported by the UK Engineering and Physical Sciences Research Council (EPSRC), under Research Grant No. EP/V002821/1.
C.T. acknowledges support by the Cluster of Excellence ``Precision Physics, Fundamental Interactions, and
Structure of Matter'' (PRISMA$^+$ EXC 2118/1) funded by the Deutsche Forschungsgemeinschaft (DFG, German Research
Foundation) within the German Excellence Strategy (Project No. 390831469).

\newpage

\begin{appendix}
\begin{section}{Path integral over the gauge-fixed field configuration space}
\label{Appendix:path:integral}

In this appendix, we derive the path integral formulation for the partition function~\eqref{eq:partition_function} on $T^4$ in the constrained configuration space ${\cal A}$. The partition funciton is defined as a trace over the physical Hilbert space,
\begin{align}
 Z=&\sum_a \langle\Psi^{(a)}|  \e^{-\beta H}|\Psi^{(a)}\rangle\notag\\
 %%%%
 =&\sum_n\int {\cal D}{\bf A}_1({\bf x})\int {\cal D}{\bf A}_2({\bf x})
\langle\Psi^{(a)}|{\bf A}_1({\bf x})\rangle\langle {\bf A}_1({\bf x})|  \e^{-\beta H}|{\bf A}_2({\bf x})\rangle\langle {\bf A}_2({\bf x})|\Psi^{(a)}\rangle\,.
\label{eq:Z1}
\end{align}
Above, $|\Psi^{(a)}\rangle$ represent the physical states in the constrained Hilbert space, and we have inserted the spectral resolution of the identity in the full unconstrained space in terms of projectors into eigenvectors $|{\bf A}(\bf x)\rangle$ of the field operators in the Heisenberg picture,
\begin{align}\label{eq:Iphys}
    \mathds{1}_{\rm phys}= \int {\cal D}{{\bf A}({\bf x})}|{\bf A}({\bf x})\rangle \langle {\bf A}({\bf x})|P_{\rm phys}\,.
\end{align}
In the previous equation, $P_{\rm phys}$ is a projector into the space of physical states $|\Psi^{(a)}\rangle$. In particular,  $P_{\rm phys}|\Psi^{(a)}\rangle = |\Psi^{(a)}\rangle$, which allows us to drop the projectors in Eq.~\eqref{eq:Z1}.
The inner product $\langle \bullet |\bullet\rangle$ is defined over the full unconstrained field space, and so it also integrates over all gauge-equivalent configurations. We can identify the inner products with our wave-functionals, 
\begin{align}\label{eq:wave_fun_kets}
 \langle {\bf A}({\bf x})|{\Psi}^{(a)}\rangle= {\Psi}^{(a)}[{\bf A}]= \e^{{\rm i}(\theta-\theta_\Pi)W[{\bf A}]}{\Psi'
}^{(a)}[{\bf A}]=\e^{{\rm i}(\theta-\theta_\Pi)W[{\bf A}]} {\Psi'}^{(a)}_{\rm g.i.}[{\bf A}]\,,
\end{align}
where we have used Eqs.~\eqref{Psi:prime} and \eqref{eq:Psi_prime_gi}. From Eq.~\eqref{eq:wave_fun_kets}, and using
\begin{align}
 &\e^{-{\rm i}(\theta-\theta_\Pi)W[{\bf A}_1]}\langle {\bf A}_1({\bf x})|\e^{-\beta H}|{\bf A}_2({\bf x})\rangle \e^{{\rm i}(\theta-\theta_\Pi)W[{\bf A}_2]}\notag\\
 %%%
 =&\langle {\bf A}_1({\bf x})|\e^{-{\rm i}(\theta-\theta_\Pi)W[{\bf A}]}\e^{-\beta H}\e^{{\rm i}(\theta-\theta_\Pi)W[{\bf A}]}|{\bf A}_2({\bf x})\rangle=\langle {\bf A}_1({\bf x})|\e^{-\beta {H'}}|{\bf A}_2({\bf x})\rangle
\end{align}
we obtain
\begin{align}\label{eq:Z_torus}\begin{aligned}
 &Z=
 %%%%
 &\sum_a\int {\cal D}{\bf A}_1({\bf x})\int {\cal D}{\bf A}_2({\bf x}){\Psi'}^{(a)*}_{\rm g.i.}[{\bf A}_1]{\Psi'}^{(a)}_{\rm g.i.}[{\bf A}_2]
\langle {\bf A}_1({\bf x})|\e^{-\beta {H'}}|{\bf A}_2({\bf x})\rangle\,.
\end{aligned}\end{align}
One can recognize  $\sum_a {\Psi'}^{(a)*}_{\rm g.i.}[{\bf A}_1]{\Psi'}^{(a)}_{\rm g.i.}[{\bf A}_2]$ as a projector into the physical eigenstates in the diagonal basis, 
\begin{align}\label{eq:spectral_P_phys}
 \sum_a {\Psi'}^{(a)*}_{\rm g.i.}[{\bf A}_1]{\Psi'}^{(a)}_{\rm g.i.}[{\bf A}_2]=P'_{\rm phys}({\bf A}_1,{\bf A}_2)\,.
\end{align}

As seen in Section~\ref{sec:quantum:constraints}, in the diagonal basis the physical eigenstates satisfy Eq.~\eqref{inv:Psi:prime}, i.e. they are gauge invariant. As a consequence of this, the physical wave functions can be taken as having support on the hypersurface $\mathcal{A}$, which can parameterized  by the field directions $\delta {\bf A}(\sigma_\parallel)$ projected on $\A$. This leads to a well-defined inner product~\eqref{inner:product:g.f.}.
Note that the restriction of the physical wave functionals to $\mathcal{A}$ is only really possible in the diagonal basis, as in an ordinary basis the wave functional still changes under gauge transformations. 
Let the hypersurface $\mathcal{A}$ be defined piecewise in different regions of field space by equations of the form
\begin{align}
F({\bf A})=0\,.    
\end{align}
By assumption, $\mathcal{A}$  intersects each of the orbits associated with gauge transformations only once. This implies that, given an arbitrary field configuration ${\bf A}$, there exists a unique gauge transformation $\hat U({\bf A})$ such that ${\bf A}_{\hat U(\bf A)}\equiv {\bf A}_\mathcal{A} \in\mathcal{A}$, or equivalently $F({\bf A}_{\hat U(\bf A)})=0$. 
The physical projector in Eq.~\eqref{eq:spectral_P_phys} can be taken as 
\begin{align}\label{eq:Pprimephys}
    P'_{\rm phys}({\bf A}_1,{\bf A}_2)\propto\delta(\bf A_1-\bf{A_1}_\mathcal{A})\delta ({\bf A}_1-{\bf A}_2)\,
\end{align}
as this operator has zero support outside $\mathcal{A}$, and is proportional to the identity inside it. Then Eq.~\eqref{eq:Z_torus} finally becomes 
\begin{align}
\label{eq:Z-torus4}
    Z\propto  \int_{\A} {\cal D}{{\bf A}}_\A ({\bf x})\,
\langle {{\bf A}}_\A ({\bf x})|\e^{-\beta {H'}}|{{\bf A}}_{\A}({\bf x})\rangle\,.
\end{align}
The factor $\langle {\bf A}_{\mathcal A}({\bf x})|\e^{-\beta {H'}}|{\bf A}_{\mathcal A}({\bf x})\rangle$ is equivalent to an ordinary path integral with paths going to a fixed configuration ${\bf A}_\mathcal{A}({\bf x})$ at $\tau=0$ and $\tau=\beta$. As a consequence, $Z$ is now given by a path integration over periodic boundary conditions, where the latter are to be inside the gauge-fixed surface $\cal A$.

The path integral~(\ref{eq:Z-torus4}) can now be brought to a gauge-fixed form, as can be seen when  $\langle {\bf A}_{\mathcal A}({\bf x})|\e^{-\beta {H'}}|{\bf A}_{\mathcal A}({\bf x})\rangle$ is written in terms of a product of amplitudes over infinitesimal time intervals. To see this, consider a partition of the time interval $0\leq \tau\leq \beta$ into $N+1$ segments $\tau_i=i\beta/N,i=0,\dots,N$,  such that for each intermediate time we insert two spectral resolutions of the identity \eqref{eq:Iphys} in the physical Hilbert space. This gives 
\begin{align}\label{eq:Amplitude0}\begin{aligned}
&\langle {\bf A}_{\mathcal A}({\bf x})|\e^{-\beta {H'}}|{\bf A}_{\mathcal A}({\bf x})\rangle \\
&\propto \prod_{i=1}^{N-1}  \int {\cal D}{{\bf A}}^{
    t_i}{\cal D}{\tilde{\bf A}}^{
    t_i}\langle {\bf A}^{t_i}({\bf x})|P'_{\rm phys}|\tilde {
    \bf A}^{t_i}({\bf x})\rangle \langle\tilde {\bf A}^{t_i}({\bf x})|\e^{-\Delta \tau H'}|{\bf A}^{t_{i-1}}({\bf x})\rangle\,.
\end{aligned}\end{align}
We can insert factors of unity \`a la Faddeev--Popov for each time $t_i$, 
\begin{align}
\label{eq:FP-identity}
    \mathds{1}=\int \D U^{t_i}({\bf x})\, \delta (U^{t_i}-\hat U({\bf A}^{t_i}))=\int \D U^{t_i}({\bf x})\, \delta (F({\bf A}^{t_i}_{U^{t_i}})) \left|\det\left(\frac{\delta F({\bf A}^{t_i}_{{U^{t_i}}})}{\delta {U^{t_i}}}\right)\right|_{F=0} \,.
\end{align}
The Faddeev--Popov functional determinant depends on the specific form of $F$, but not on $U^{t_i}.$ Because of this, to avoid confusion, we will simply denote the Faddeev--Popov functional determinant as $\det_{\rm FP}({\bf A}^{t_i})$. Substituting Eqs.~\eqref{eq:FP-identity} and \eqref{eq:Pprimephys}---with ${\bf A}^{t_i}_\A ={\bf A}^{t_i}_{\hat U({\bf A}^{t_i})}$---into Eq.~\eqref{eq:Amplitude0}, one gets 
\begin{align}
    \langle {\bf A}_{\mathcal A}({\bf x})|\e^{-\beta {H'}}|{\bf A}_{\mathcal A}({\bf x})\rangle \propto & \prod_{i=1}^{N-1} \int \D {\bf A}^{t_i}\D \tilde{\bf A}^{t_i}\D U^{t_i}({\bf x})\,\delta (U^{t_i}-\hat U({\bf A}^{t_i}))\notag\\
    %%%
    \times&\delta({\bf A}^{t_i}-{\bf A}^{t_i}_{\hat U({\bf A}^{t_i})})\delta({\bf A}^{t_i}-\tilde{\bf A}^{t_i})\langle\tilde {\bf A}^{t_i}({\bf x})|\e^{-\Delta \tau H'}|{\bf A}^{t_{i-1}}({\bf x})\rangle\,.
    \end{align}
 Doing the integrations over $\tilde{\bf A}^{t_i}$ and using the constraint $\delta(U^{t_i}-\hat U({\bf A}^{t_i}))$ in $\delta({\bf A}^{t_i}-{\bf A}^{t_i}_{\hat U({\bf A}^{t_i})})$ gives
    \begin{align}
    %%%%
      & \langle {\bf A}_{\mathcal A}({\bf x})|\e^{-\beta {H'}}|{\bf A}_{\mathcal A}({\bf x})\rangle \notag\\
       %%%
       \propto&\prod_{i=1}^{N-1}  \int\D {\bf A}^{t_i} \D U^{t_i}({\bf x})\,
       \delta (F({\bf A}^{t_i}_{U^{t_i}})) |{\det}_{\rm FP}({\bf A}^{t_i})|
       \delta({\bf A}^{t_i}-{\bf A}^{t_i}_{ U^{ t_i}})\langle {\bf A}^{t_i}({\bf x})|\e^{-\Delta \tau H'}|{\bf A}^{t_{i-1}}({\bf x})\rangle\notag\\
       =&\prod_{i=1}^{N-1}  \int\D {\bf A}^{t_i} \D U^{t_i}({\bf x})\,
       \delta (F({\bf A}^{t_i})) |{\det}_{\rm FP}({\bf A}^{t_i})|
       \delta({\bf A}^{t_i}-{\bf A}^{t_i}_{ U^{ t_i}})\langle {\bf A}^{t_i}({\bf x})|\e^{-\Delta \tau H'}|{\bf A}^{t_{i-1}}({\bf x})\rangle\notag\\
       \propto&\prod_{i=1}^{N-1}\int\D {\bf A}^{t_i}\, |{\det}_{\rm FP}({\bf A}^{t_i})|\,\delta(F({\bf A}^{t_i}))\langle {\bf A}^{t_i}({\bf x})|\e^{-\Delta \tau H'}|{\bf A}^{t_{i-1}}({\bf x})\rangle\,.
       \label{eq:Amplitudefinal}
\end{align}
In the first step we have used Eq.~\eqref{eq:FP-identity}. The product of factors of determinants and $\delta$-functions  at fixed times gives rise in the limit of $N\rightarrow \infty$ to the usual Faddeev--Popov determinant and overall $\delta$-function, this time understood as functionals of field configurations that depend on both space and time. Therefore, at all times the gauge field configurations to be integrated are constrained on $\A$.

The resulting path integral still only depends on the spatial fields ${\bf A}$. To recover the familiar path integration over trajectories $A_\mu(t, \vec{x})$, with Lorentz invariance made manifest, one can make use of the fact that the states in the physical Hilbert space automatically satisfy Gau{\ss}' law, as discussed in Section~\ref{sec:Gauss}. Hence, one can insert the unity operator in the constrained space as a projector $P_{\rm Gau{\ss}}$ into field configurations that satisfy  Gau{\ss}' law. Up to infinite, gauge-invariant normalization constants, one can write
\begin{align}
    P_{\rm Gau{\ss}}|{\bf A}(\bf x)\rangle \propto \delta({\bf D}\cdot{\bf E)}|{\bf A}(\bf x)\rangle\,.
\end{align}
As shown in Ref.~\cite{Fradkin:2021zbi}, the factors of $P_{\rm Gau\text{\ss}}$ can be traded for integrations over the temporal components $A_0$ of the gauge potential, leading to a path integral over all field components $A_\mu$. The integrand ends up being of the usual form $\e^{-S_{\rm E}-S_{\rm g.f.}-S_{\rm E,ghost}}$, with $S_{\rm E}$ the ordinary Euclidean gauge-invariant action associated with the Hamiltonian $H'$. Thus, we can write
\begin{align}\begin{aligned}
    %%%%
       \langle {\bf A}_{\mathcal A}({\bf x})|\e^{-\beta {H'}}|{\bf A}_{\mathcal A}({\bf x})\rangle=\int_{ {\bf A}_{\mathcal A}({\bf x})} \D{ A}(t,{\bf x}) \e^{-S'_{\rm E}-S_{\rm g.f.}-S_{\rm E,ghost}}\,,
\end{aligned}\end{align}
where the path integral has periodic boundary conditions, with the fields going to the configuration  ${\bf A}_{\mathcal A}({\bf x})$ at  $\tau=0,\beta$. Above, to avoid cluttering the notation, we have suppressed the integral over the ghost fields. The full partition function \eqref{eq:Z-torus4} becomes 
\begin{align}\begin{aligned}
    %%%%
       Z=\int _\A \D {\bf A}_\A({\bf x})\int_{ {\bf A}_{\mathcal A}({\bf x})} \D{ A}(t,{\bf x})\e^{-S'_{\rm E}-S_{\rm g.f.}-S_{\rm E,ghost}}=\int_{\A;\rm  periodic} \D{ A}(t,{\bf x})\e^{-S'_{\rm E}-S_{\rm g.f.}-S_{\rm E,ghost}}\,.
\end{aligned}\end{align}
where now one has to consider generic periodic field configurations with no temporal component at  $\tau=0,\beta$.  Note that this does not imply that there is no sum over topological sectors. Since our present gauge condition is that the vector potentials are in the hypersurface ${\cal A}$, this means that identical physical configurations are represented by the same $\mathbf A$. So the sum of the topological flux through the spatial volumes at $\tau=0$ and $\tau=\beta$ is zero. However, topological flux can still permeate through the timelike surfaces of the torus, and per the reasoning given in Section~\ref{sec:topology:T4}, the topological charge  remains quantized on the four-torus. Upon transforming the gauge fixing, we can eventually choose gauges so that the topological flux goes through the spacelike surfaces. Or alternatively in the gauge-fixed form, if desired, one may even choose different hypersurfaces ${\cal A}$ and ${\cal A}'$ which differ from each other by a gauge transformation at $\tau=0$ and $\tau=\beta$, respectively.
Separating explicitly the contributions from the different topological sectors, the result can be written as in Eq.~\eqref{eq:Ztorusfinal}.

\end{section}
\end{appendix}

\bibliographystyle{utphys}
\bibliography{torus.bib}{}

\end{document}